\documentclass[twocolumn,trackchanges]{aastex631}

\usepackage{amsmath}
\usepackage{lipsum}
\usepackage{graphics}
\usepackage{CJK}

\shorttitle{SMBH-Galaxy Mass Ratio at z$<$4}
\shortauthors{Sun et al.}

\usepackage{graphicx}

\begin{document}

\title{No evidence for a significant evolution of $M_{\bullet}$-$M_*$ relation in massive galaxies up to z$\sim$4}

\correspondingauthor{Yang Sun}
\email{sunyang@arizona.edu}

\author[0000-0001-6561-9443]{Yang Sun}
\affiliation{Steward Observatory, University of Arizona,
933 North Cherry Avenue, Tucson, AZ 85719, USA}

\author[0000-0002-6221-1829]{Jianwei Lyu (\begin{CJK}{UTF8}{gbsn}吕建伟\end{CJK})}
\affiliation{Steward Observatory, University of Arizona,
933 North Cherry Avenue, Tucson, AZ 85719, USA}

\author[0000-0003-2303-6519]{George H. Rieke}
\affiliation{Steward Observatory, University of Arizona,
933 North Cherry Avenue, Tucson, AZ 85719, USA}

\author[0000-0001-7673-2257]{Zhiyuan Ji}
\affiliation{Steward Observatory, University of Arizona,
933 North Cherry Avenue, Tucson, AZ 85719, USA}

\author[0000-0002-4622-6617]{Fengwu Sun}
\affiliation{Center for Astrophysics $|$ Harvard \& Smithsonian, 60 Garden St., Cambridge MA 02138 USA}

\author[0000-0003-3307-7525]{Yongda Zhu}
\affiliation{Steward Observatory, University of Arizona,
933 North Cherry Avenue, Tucson, AZ 85719, USA}

\author[0000-0002-8651-9879]{Andrew J. Bunker}
\affiliation{Department of Physics, University of Oxford, Denys Wilkinson Building, Keble Road, Oxford OX1 3RH, UK}

\author[0000-0002-1617-8917]{Phillip A. Cargile}
\affiliation{Center for Astrophysics $|$ Harvard \& Smithsonian, 60 Garden St., Cambridge MA 02138 USA}

\author[0000-0001-8522-9434]{Chiara Circosta}
\affiliation{European Space Agency (ESA), European Space Astronomy Centre (ESAC), Camino Bajo del Castillo s/n, 28692 Villanueva de la Ca \~{n}ada, Madrid, Spain}
\affiliation{Department of Physics \& Astronomy, University College London, Gower Street, London, WC1E 6BT, UK} 

\author[0000-0003-2388-8172]{Francesco D'Eugenio}
\affiliation{Kavli Institute for Cosmology, University of Cambridge, Madingley Road, Cambridge, CB3 0HA, UK}
\affiliation{Cavendish Laboratory -- Astrophysics Group, University of Cambridge, 19 JJ Thomson Avenue, Cambridge, CB3 0HE, UK}
\affiliation{INAF -- Osservatorio Astronomico di Brera, via Brera 28, I-20121 Milano, Italy}

\author[0000-0003-1344-9475]{Eiichi Egami}
\affiliation{Steward Observatory, University of Arizona, 933 North Cherry Avenue, Tucson, AZ 85719, USA}

\author[0000-0003-4565-8239]{Kevin Hainline}
\affiliation{Steward Observatory, University of Arizona, 933 North Cherry Avenue, Tucson, AZ 85719, USA}

\author[0000-0003-4337-6211]{Jakob M. Helton}
\affiliation{Steward Observatory, University of Arizona, 933 North Cherry Avenue, Tucson, AZ 85719, USA}


\author[0000-0002-5104-8245]{Pierluigi Rinaldi}
\affiliation{Steward Observatory, University of Arizona, 933 North Cherry Avenue, Tucson, AZ 85719, USA}

\author[0000-0002-4271-0364]{Brant E. Robertson}
\affiliation{Department of Astronomy and Astrophysics, University of California, Santa Cruz, 1156 High Street, Santa Cruz, CA 95064, USA}

\author{Jan Scholtz}
\affiliation{Kavli Institute for Cosmology, University of Cambridge, Madingley Road, Cambridge, CB3 0HA, UK} 
\affiliation{Cavendish Laboratory -- Astrophysics Group, University of Cambridge, 19 JJ Thomson Avenue, Cambridge, CB3 0HE, UK}

\author[0000-0003-4702-7561]{Irene Shivaei} 
\affiliation{Centro de Astrobiolog\'ia (CAB), CSIC-INTA, Ctra. de Ajalvir km 4, Torrej\'on de Ardoz, E-28850, Madrid, Spain}

\author[0000-0002-9720-3255]{Meredith A. Stone}
\affiliation{Steward Observatory, University of Arizona,
933 North Cherry Avenue, Tucson, AZ 85719, USA}

\author[0000-0002-8224-4505]{Sandro Tacchella}
\affiliation{Kavli Institute for Cosmology, University of Cambridge, Madingley Road, Cambridge, CB3 0HA, UK}
\affiliation{Cavendish Laboratory -- Astrophysics Group, University of Cambridge, 19 JJ Thomson Avenue, Cambridge, CB3 0HE, UK}

\author[0000-0003-2919-7495]{Christina C. Williams}
\affiliation{NSF's National Optical-Infrared Astronomy Research Laboratory, 950 North Cherry Avenue, Tucson, AZ 85719, USA}
\affiliation{Steward Observatory, University of Arizona, 933 North Cherry Avenue, Tucson, AZ 85719, USA}

\author[0000-0001-9262-9997]{Christopher N. A. Willmer}
\affiliation{Steward Observatory, University of Arizona, 933 North Cherry Avenue, Tucson, AZ 85719, USA}

\author[0000-0002-4201-7367]{Chris Willott}
\affiliation{NRC Herzberg, 5071 West Saanich Rd, Victoria, BC V9E 2E7, Canada}



\begin{abstract}

Over the past two decades, tight correlations between black hole masses ($M_\bullet$) and their host galaxy properties have been firmly established for massive galaxies ($\log(M_*/M_{\odot})\gtrsim10$) at low-$z$ ($z<1$), indicating coevolution of supermassive black holes and galaxies. However, the situation at high-$z$, especially beyond cosmic noon ($z\gtrsim2.5$), is controversial. With a combination of \emph{JWST} NIRCam/wide field slitless spectroscopy (WFSS) from FRESCO, CONGRESS and deep multi-band NIRCam/image data from JADES in the GOODS fields, we study the black hole to galaxy mass relation at z$\sim$1--4. After identifying 18 broad-line active galactic nuclei (BL AGNs) at $1<z<4$ (with 8 at $z>2.5$) from the WFSS data, we measure their black hole masses based on broad near-infrared lines (Pa $\alpha$, Pa $\beta$, and He\,I $\lambda$10833\,\AA), and constrain their stellar masses ($M_{*}$) from AGN-galaxy image decomposition or SED decomposition. Taking account of the observational biases, the intrinsic scatter of the $M_{\bullet}-M_{*}$ relation, and the errors in mass measurements, we find no significant difference in the $M_{\bullet}/M_{*}$ ratio for 2.5 $< $ z $ <$ 4 compared to that at lower redshifts ($1 < z < 2.5$), suggesting no evolution of the $M_{\bullet} - M_{*}$ relation at $\log(M_*/M_{\odot})\gtrsim10$ up to z$\sim$4.

\end{abstract}

\keywords{
\href{http://astrothesaurus.org/uat/16}{Active galactic nuclei (16)},
\href{http://astrothesaurus.org/uat/98}
{Supermassive black holes (98)},
\href{http://astrothesaurus.org/uat/573}{Active galaxies (573)},
\href{http://astrothesaurus.org/uat/594}{Galaxial evolution (594)}
}


\section{Introduction} \label{sec:intro}

With the discovery of quasars \citep{Hazard1963,Schmidt1963}, accretion onto supermassive black holes (SMBHs) has been appreciated as the second major source of electromagnetic radiation in the Universe, next to stellar radiation. The argument that the evolution of these two fundamental energy sources might be linked is highly influential in modern astronomy and has fostered numerous investigations on how the correlation between SMBHs and their hosts is established and maintained (see reviews by e.g., \citealt{Alexander2012, Kormendy2013, Heckman2014,Harrison2017}). The potential scenarios range from a direct causal connection such as feedback due to winds and outflows launched by the active galactic nuclei (AGN) that regulate the growth of the host galaxy \citep[e.g.,][]{Springel2005, Hopkins2008b, Hopkins2008, Fabian2012}, to a simple consequence of the growth of galaxies through merging without a physical coupling between the galaxy and black hole growth \citep[e.g.,][]{Peng2007, Jahnke2011}. 

Insights into this relationship and the underlying physical mechanisms can be obtained from how it evolves with cosmic time. In the low-$z$ Universe, the masses of SMBHs ($M_\bullet$) have been firmly established to correlate with many properties of their hosts, notably the mass ($M_{\rm b}$) of the spheroid component; e.g., \citealt{Magorrian1998, Haring2004, Kormendy2013}) and its velocity dispersion ($\sigma_{\rm b}$; \citealt{Ferrarese2000, Gebhardt2000, Tremaine2002}). At high-$z$, due to various observational limitations, the explorations have been largely focused on the ratio between the SMBH masses and total stellar masses ($M_{*}$)  $M_\bullet$/$M_{*}$  of the quasar population up to z $\sim$ 2--2.5, and the results were controversial. At the massive regime ($\log(M_{*}/M_{\odot})>10$), several observational works claimed that galaxies at z $\sim$ 2 tend to have higher $M_\bullet$/$M_{*}$ than the local value \citep[e.g.,][]{Peng2006,Merloni2010, Zhang2023}, suggesting a significant evolution of this mass scaling relation over the past $\sim$ 9 Gyr; Meanwhile, many other studies \citep[e.g.,][]{Schramm2013,Mechtley2016,Ding2020,Suh2020,JLi2023, Mountrichas2023} found no significant evolution of $M_\bullet$/$M_{*}$. The lower-mass regime ($\log(M_{*}/M_{\odot})<9.5$) is not well explored in the standard scaling relations \citep{Reines2015, Greene2020}. A few dwarf galaxies with broad line AGNs at $0.4<z<3$ have been reported recently, and they overall have higher $M_\bullet$/$M_{*}$ than the local relation \citep{Mezcua2023,Mezcua2024}, but this result has not yet been integrated into the broader context of low mass galaxies overall.

A partial explanation for these discrepancies is that there are multiple measurement biases affecting the $M_\bullet$/$M_{*}$ determinations at high redshift. For example, the single-epoch SMBH masses are derived from ultraviolet lines of CIV and MgII whose widths can be increased by outflows and other effects \citep[e.g.,][]{Shen2012, Le2020, Zuo2020}. The AGNs most readily observed will be the most luminous and hence will tend to be overmassive relative to typical behavior, resulting in a bias toward high values of $M_\bullet$/$M_{*}$\citep[e.g.,][]{Willott2005, Lauer2007}, which is the well known ``Lauer bias''. 
In fact, \citet{Schulze2011,Schulze2014} tested this possibility and found that the differences in a number of studies in the slope of the $M_\bullet$/$M_{*}$ relation at high and low redshift disappears when corrected for the ``Lauer bias''. Taking the selection bias into account, it seems a consensus is emerging that there is little evolution of $M_\bullet$/$M_{*}$ up to $z \sim 2.5$ \citep{Sun2015, Suh2020, JLi2023, Mountrichas2023, Tanaka2024}, although this issue is perhaps not fully settled \citep[e.g.,][]{Zhang2023}.

With the successful launch and operation of \emph{JWST}, the study of the $M_{\bullet}-M_{*}$ relation has been pushed to much higher redshifts. Quasars at $z\sim6$ with direct host stellar emission constraints from Near Infrared Camera (NIRCam; \citealt{Rieke2023}) appear to show relatively large values of $M_\bullet$/$M_{*}$ ($\log (M_\bullet$/$M_{*}) \sim -1$) compared to their lower-redshift counterparts ($\log (M_\bullet$/$M_{*}) \sim -2.5$) \citep{Ding2023,Yue2024,Stone2024}. Significant evolution in this relation has also been suggested for relatively less massive SMBHs in Seyfert luminosity AGNs at $4 < z < 7$ \citep{Ubler2023,Maiolino2023,Harikane2023,Pacucci2023}. These results reveal a sharp contrast with those at $z\lesssim2.5$, resulting in the need to fill in the redshift gap between $z\sim2.5$ and $z\sim4$ to establish a complete picture on the evolution of $M_{\bullet}-M_{*}$ relation across the cosmic time.

In this work, we will present a comprehensive study on the $M_{\bullet}$--$M_{*}$ relation in massive galaxies ($\log(M_{*}/M_{\odot})>10$) at $z\sim$1--4 based on a powerful combination of 
the NIRCam/Wide-Field Slitless Spectroscopy (WFSS) surveys FRESCO \citep[PID: 1895,][]{Oesch2023} and CONGRESS (PID: 3577, Sun et al. in prep.) and deep multi-band NIRCam images from JADES \citep{Eisenstein2023} in GOODS-S and GOODS-N fields \citep{Giavalisco2004}. In contrast to previous ground-based $z\sim$ 1--2 studies limited to rest-frame UV to optical lines, we are able to accurately constrain the black hole masses from the near-infrared (NIR) broad emission lines (Paschen lines and He\,I $\lambda$10833\,\AA), which are much less affected by dust extinction and galaxy contamination. Moreover, the superior spatial resolution and sensitivity of multi-band NIRCam images enable robust AGN-galaxy decomposition with a wide range of wavelength coverage, allowing accurate measurements of host stellar masses from SED fittings on the AGN-subtracted galaxy emission. Finally, the survey nature of NIRCam/WFSS data as well as the deep multi-wavelength coverage from X-ray to the radio data in the GOODS fields provide the peerless opportunity to build a complete sample and understand the selection biases.

This paper is organized as follows: We introduce the AGN sample and the relevant \emph{JWST} data for this project in Section 2. Section 3 describes how the black hole and galaxy masses of our sample are measured. In Section 4, we present the measurements of the black hole to galaxy mass ratios at $z\sim$ 1--4 and analyze these results on a basis consistent with the approach used in the lower-redshift studies and evaluate measurement biases with a Monte Carlo method. We discuss the possible implications of our results for studies at higher redshift in Section 5 and conclude this work in Section 6. 

Throughout this paper, we assume a standard $\Lambda$CDM universe with cosmological parameters $\mathrm{H_0} =
70~\mathrm{km~s^{-1}~Mpc^{-1}}$, $\Omega_{\Lambda} = 0.7$, and $\Omega_{\mathrm{m}} = 0.3$.

\section{Data and Sample Selection} \label{sec:data}

\subsection{Spectroscopic Data} \label{sec:fresco&congress}

To search for AGN broad line emission, we used the spectroscopic data obtained with JWST/NIRCam Wide Field Slitless Spectroscopy (WFSS) from the FRESCO survey \citep{Oesch2023} in the F444W band ($\lambda\sim$ 3.9--5.0~$\mu m$) and in both GOODS-S and GOODS-N, and the CONGRESS survey (Sun et al, in prep) in the F356W ($\lambda\sim $3.1--4~$\mu m$) band in GOODS-N only. The former survey covers a 7.4'$\times$8.4' area in both GOODS fields with the row-direction grisms on both modules of JWST/NIRCam providing a spectral resolution of $\sim$1590--1680 from 3.9 to 5.0~$\mu m$. The CONGRESS program covers the GOODS-N field with a nearly identical footprint to FRESCO and with a spectral resolution of $\sim$1400--1610 from 3.1 to 4.0~$\mu m$. The two programs reach similar line sensitivities of $2\times10^{-18}~{\rm erg~s^{-1}cm^{-2}}$.


All these NIRCam/WFSS data were processed by the publicly available reduction routine presented in \citet{Sun2023}\footnote{\url{https://zenodo.org/records/14052875} \citep{https://doi.org/10.5281/zenodo.14052875}}. 
We first processed the NIRCam data through the standard JWST stage-1 calibration pipeline\footnote{\url{https://zenodo.org/records/8140011} \citep{https://doi.org/10.5281/zenodo.6984365}} \texttt{v1.11.2}.
For each individual grism exposure, we assigned world coordinate system (WCS) information, performed flat-fielding, and removed the $\sigma$-clipped median sky background. 
The WCS of the grism exposures is registered to Gaia DR3 \citep{Gaia2023} using the NIRCam short-wavelength imaging data taken at the same time.
For each of our targets, we extracted the 2D spectra from individual grism exposures, and coadded them in a common wavelength (1 nm/pixel) and spatial (0\farcs06/pixel) grid.
Prior to our scientific spectral extraction, we also extracted spectra of bright point sources ($\lesssim$\,21\,AB mag) to assure the accuracy of spectral tracing function and spectral flux calibration. 
We also extracted the spectra of galaxies with known ground-based spectroscopic redshifts, measuring the line center of detected Paschen $\alpha$ and $\beta$ lines to ensure the wavelength calibration error at $<$\,1\,nm.
We then optimally extracted the 1D spectra of our targets from coadded 2D spectra using their surface brightness profile in the F444W band \citep{Horne1986}.



\subsection{Imaging Data} \label{sec:jades}


Multi-band NIRCam images with superior spatial resolution and sensitivity are available for both the GOODS-S and GOODS-N fields from JADES \citep{Eisenstein2023}. We used these images to conduct AGN-galaxy morphology decomposition and stellar mass estimation.  The total overlapping area between the JADES and FRESCO footprints is approximately 35 square arcminutes in GOODS-N and about 46 square arcminutes in GOODS-S. JADES also overlaps with CONGRESS in GOODS-N in a similar area since CONGRESS covers almost the same area as FRESCO there. Seven NIRCam broad-band images were selected to cover a wide range of wavelengths: F090W, F115W, F150W, F200W, F277W, F356W, and F444W. For galaxies at z $\sim$ 1--3.5, these NIRCam filters nicely cover the rest optical to near-IR spectral energy distributions (SEDs) of the galaxies, allowing reasonable constraints on the galaxy stellar properties. The angular resolutions of these images range from 0.030\arcsec\ to 0.145\arcsec, corresponding to physical scales of $\sim$ 0.2--1.2 kpc for galaxies at $z\sim1$--3.5. 


The four BL AGNs among our sample\footnote{BL AGN selection will be described in Section~\ref{sec:AGNselect}.} within the FRESCO area but outside the JADES footprint do not have full NIRCam wide-band data, but do have NIRCam F182M, F210M and F444W data from the FRESCO survey. In addition, deep multi-band images in the optical to the near-IR are also available from HST/ACS and HST/WFC3 \citep[e.g.,][]{Koekemoer2011}. However, given the limited NIRCam bands as well as the different spatial resolutions of JWST and HST in the near-IR, we decided not to conduct image decompositions for those four targets but to carry out SED decompositions of the integrated photometry (see details in Section~\ref{sec:sedfitting}).

\subsection{BL AGN Sample Selection} \label{sec:AGNselect}
\begin{figure*}[htb]
\centering
\includegraphics[width=1\textwidth]{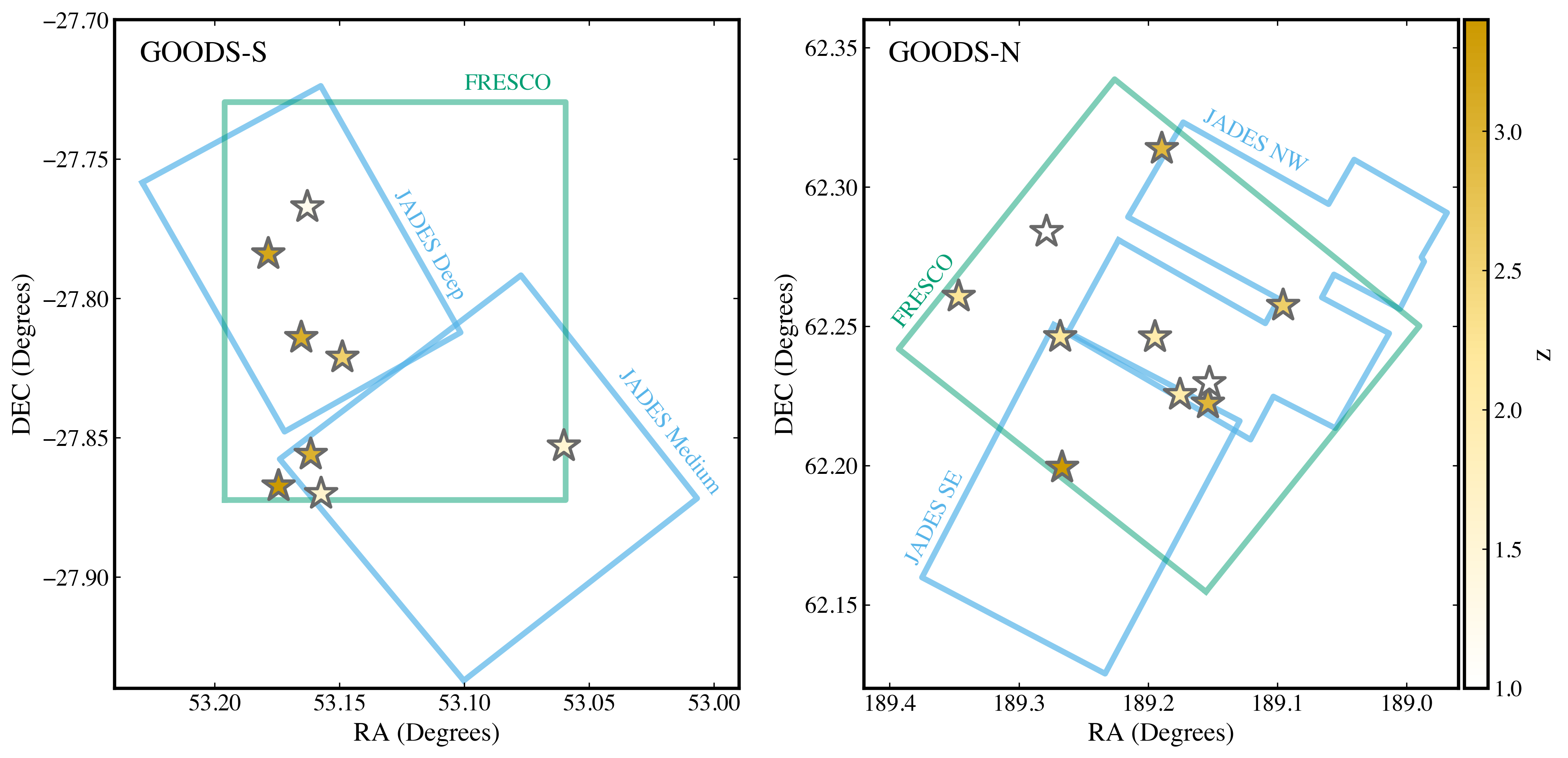}
\caption{Spatial locations of our 18 $1<z<4$ BL AGN sample (stars) in the GOODS-S (left) and GOODS-N (right) fields, color-coded by redshift. The JADES and FRESCO footprints are drawn with blue and green lines, respectively.}
\label{Fig:skyplot}
\end{figure*}

Considering the huge number of galaxies observed by FRESCO and CONGRESS and the
frequent galaxy contamination, a blind search for AGN broad-line features from
the grism data across the field is not very practical. Therefore, we built
an AGN sample with the available multi-wavelength data first and then inspected 
the corresponding NIRCam/grism spectra to identify broad-line AGNs. 
{\it This selection approach is consistent with previous studies of the mass scaling relation at z $\lesssim$ 2.5 that were based on the AGN samples selected by AGN features in multi-wavelength ranges and follow-up broad-line detection.} In fact, the GOODS-S
and GOODS-N fields have been extensively covered by ground- and space-based
telescopes with the deepest X-ray, optical, infrared, and radio data in the sky, offering the best resources to build a complete AGN sample.
With a combination of \textit{Chandra}, \textit{Hubble}, \textit{Spitzer}, and JVLA data, \citet[]{Lyu2022} have carried out a comprehensive pre-JWST search of AGN in GOODS-S and reported $\sim$900 candidates across
the field. With the newly obtained JWST JADES/NIRCam and SMILES/MIRI data \citep{Alberts2024, Rieke2024}, \citet{Lyu2024} further
improved the AGN census near the central region of GOODS-S. We combined the AGN catalogs reported in these two papers and extracted the FRESCO NIRCam F444W/grism spectra. For GOODS-N, following
the same techniques in \citet{Lyu2022}, a similar 
panchromatic pre-JWST AGN search with relatively shallower X-ray and radio data has been conducted with $\sim$700 AGN candidates revealed in the field (Lyu et al., in prep.). We started from this GOODS-N AGN sample and extracted the corresponding FRESCO and CONGRESS grism spectra in F444W and F356W.

The final sample analyzed in this work was selected by requiring a broad line detection in the FRESCO or CONGRESS spectra of the AGN sample described above. We limited the redshift range to be within 1 $<$ z $<$ 4, thus Pa $\alpha$, Pa $\beta$, or He\,I $\lambda$10833\,\AA ~are covered by the FRESCO or CONGRESS wavelength ranges. In other words, our BL AGNs must have at least one of the three broad NIR lines mentioned above in their spectra. 
For the initial sample selection, we used the redshift values on the SIMBAD website, as reported by previous works. Although most of these galaxies have well-constrained spectroscopic redshifts, some have only photometric redshifts with relatively large uncertainties. Therefore, we also remeasured spectroscopic redshifts for them during the detailed line profile fitting (see Section~\ref{sec:BHmass}). 

In the next step, to identify BL AGN candidates, we did a preliminary fit to the line profiles with a Gaussian component, selecting BL AGNs from our initial sample by requiring a broad NIR line (${\rm FWHM}>1000 ~\mathrm{km\,s^{-1}}$). This line width threshold is consistent with other BL AGN identification works using JWST NIRCam/WFSS data \citep{Matthee2023} or NIR BL features \citep{Ricci2022}. After we finalized our sample, we had eight AGNs in the GOODS-S field and eleven in the GOODS-N field. After visually inspecting their JWST/NIRCam multi-band images, we removed one of the BL AGN candidates in the GOODS-N field (GN-1030801) from our sample since it has complicated structures in its central region, which can intrinsically broaden the width of the targeted NIR lines and thus cause an overestimate of the line width of the broad component. 

Strong outflows could complicate our determination of AGN properties from line widths. For example, \citet{D'Eugenio2023} reported our BL AGN candidate GS-197911 as a post-starburst galaxy hosting an AGN and with strong ionized outflow given its blueshifted and broad [OIII] 5008$\text{\AA}$. In this case, the broad-line component of the He\,I $\lambda$10833\,\AA ~line may be contaminated by the outflow signatures. However, given that its preliminary broad line FWHM is about $3600\,\mathrm{km\,s^{-1}}$, which is much larger than the [OIII] width ($\sim1800\,\mathrm{km\,s^{-1}}$), we conclude that its broad line component is dominated by AGN broad-line region (BLR) emission and the outflow contamination would not significantly influence the black hole measurement. Thus, we still keep GS-197911 in our sample. Similarly, we rule out the possibility of outflow-dominated broadening for other BL AGN candidates 
after conducting detailed multi-component broad-line fitting (see Section~\ref{sec:BHmass}).
Therefore, we finalized our sample with 18 AGNs. 

\begin{figure}
\centering
    \includegraphics[width=1\columnwidth]{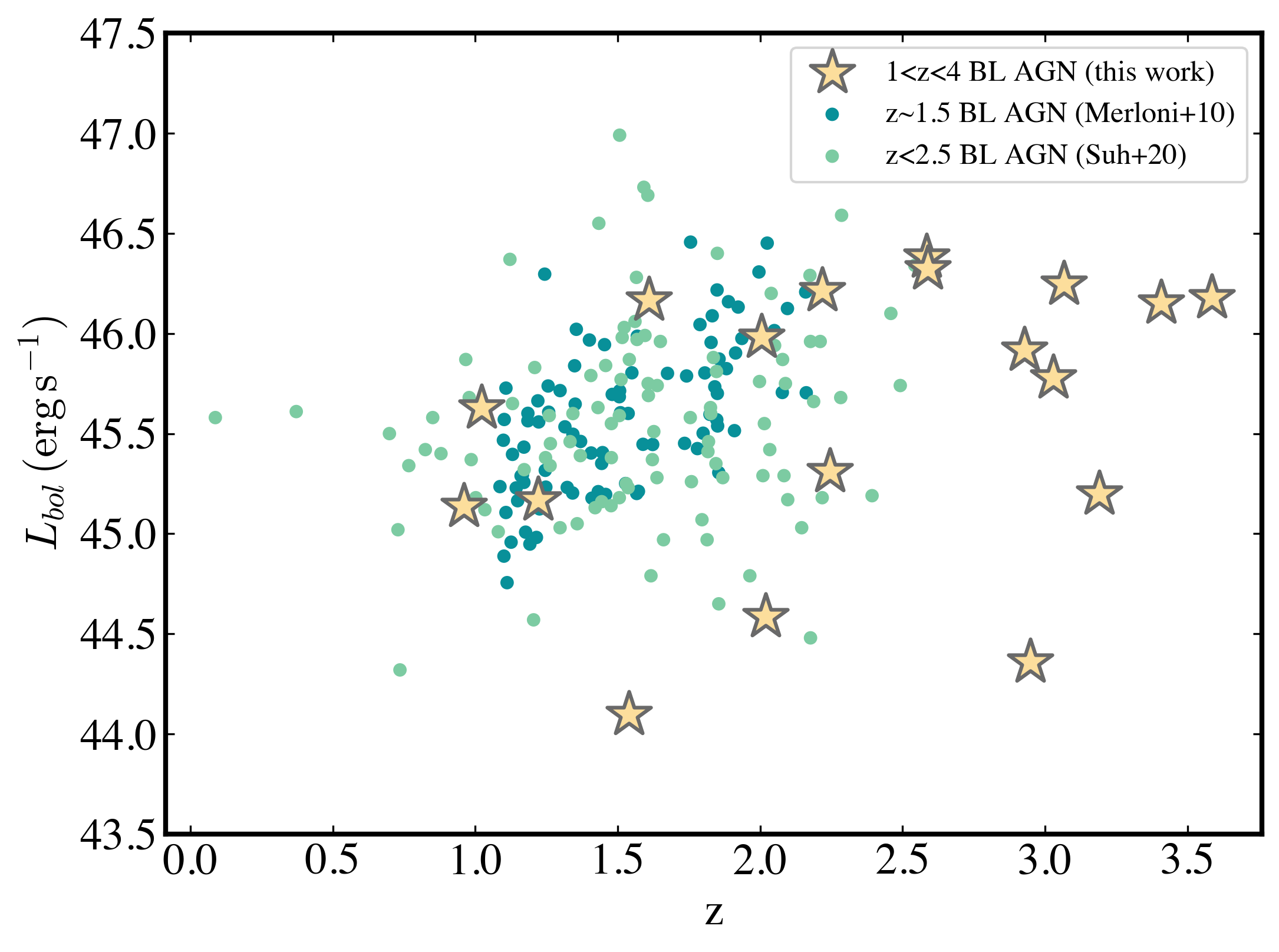}
\caption{Distribution of the sample of 18 BL AGNS  on the $L_{bol}$-$z$ plane. The bolometric luminosity of our AGNs ranges from $10^{44}$ to $10^{46.5}\,\mathrm{erg\,s^{-1}}$. Compared to the BL samples at z $\sim$ 2 from \citet{Merloni2010} (dark green) and \citet{Suh2020} (light green), our $1<z<4$ BL AGN sample is on average similar but contains examples with relatively lower $L_{bol}$ ($L_{bol}<10^{45}\,\mathrm{erg\,s^{-1}}$) and higher redshift ($z > 2.5$).}
\label{Fig:Lbol_z}
\end{figure}

The positions on the sky of our sample are shown in Figure~\ref{Fig:skyplot}, and their properties are shown in Table~\ref{Tab1}. 
The multi-wavelength studies from \citet{Lyu2022} reported that all our BL AGNs have been classified as X-ray AGNs, and the majority of them are also identified in the mid-IR as expected since they have broad near-IR emission lines. Also, their AGN continua are generally obscured in the short-wavelength range. In this case, even though all our BL AGNs have \textit{Chandra} X-ray detections \citep{Luo2017}, we used the SED-derived bolometric luminosities from \citet{Lyu2022} and converted them to get an equivalent X-ray 2-10 keV luminosity using the X-ray bolometric correction by \citet{Duras2020} for BH mass estimation (see Section~\ref{sec:BHmass}). This strategy is adopted by default as the X-ray intrinsic luminosity values reported in e.g., \cite{Luo2017} are based on simple modeling of the source X-ray flux band ratios while the SED fitting approach takes care of the obscuration in a more sophisticated way. 
We note that the SED-derived bolometric luminosities of GS-184451 and 206907 from \citet{Lyu2022}, and of GN-1000721 from Lyu et al., in prep., are quite low ($L_{bol}<10^{44}\,\mathrm{erg\,s^{-1}}$) while they do have strong X-ray emission ($L_X\sim10^{43}-10^{44}\,\mathrm{erg\,s^{-1}}$) contributed by the AGN \citep{Luo2017}. Similar X-ray bright but SED non-detected or un-identified AGNs have been report in previous work \citep{Lyu2022,Lyu2024}. After checking their SED fittings, we concluded that these three sources do not have enough mid-IR photometric data to give meaningful constraints on the AGN component. As a result, we adopted their {\it Chandra} X-ray measurements to compute X-ray 2-10 keV luminosities and the bolometric luminosities using the same bolometric correction. The final distribution of our BL AGNs on the AGN bolometric luminosity vs redshift plane is shown in Figure~\ref{Fig:Lbol_z}. Most of them help bridge the gap from previous measurements between z $\sim$ 2.5 to z = 4.

\begin{deluxetable*}{cccccccccc}
\tablecaption{Properties of the GOODS AGN with NIRCam/WFSS NIR broad-line detections\label{Tab1}}
\tablewidth{0pt}
\tabletypesize{\scriptsize}
\tablehead{ \colhead{ID$^a$}  & \colhead{RA}        & \colhead{DEC}       & \colhead{z}  & 
\colhead{$\log L_{bol}$$^b$} & \colhead{$\log L_X$$^c$} & 
\colhead{Broad Line} & \colhead{$\mathrm{FWHM_{int}}$$^d$} &
\colhead{$\log M_{\bullet}$} & \colhead{$\log M_{*}$$^e$} \\
 \colhead{}      & \colhead{(deg)}      & \colhead{(deg)}        &\colhead{} & \colhead{ ($\mathrm{erg\,s^{-1}}$)} & \colhead{($\mathrm{erg\,s^{-1}}$)} & 
 \colhead{}   & \colhead{($\mathrm{km\,s^{-1}}$)} &
 \colhead{($M_{\odot}$)} & \colhead{($M_{\odot}$)} 
}
\startdata
GS-173091 & 53.1573639 & -27.8701553 & 1.61      & 46.17               & 44.57 & Pa $\alpha$ & 2391 $\pm$ 49 & 8.07 & 10.67\\
GS-243640 & 53.174408  & -27.8674202 & 3.586  & 46.18               & 44.58 & He I & 3722 $\pm$ 671 & 8.46 &  11.22$^{*}$            \\
GS-182930 & 53.1615181 & -27.8560753 & 3.029 & 45.78               & 44.34 & He I & 1337 $\pm$ 68 & 7.45 &  10.54           \\
GS-184451 & 53.0601387 & -27.8530674 & 1.539 & 44.10               & 43.06  & Pa $\alpha$ & 1821 $\pm$ 259 & 7.10 &  10.90          \\
GS-197911 & 53.1653061 & -27.8141308 & 3.067& 46.25               & 44.62 & He I & 3470 $\pm$ 1008 & 8.49 &  11.19 \\
GS-206907 & 53.1785088 & -27.7841015 & 3.191 & 45.20               & 44.16  & He I & 5563 $\pm$ 419  & 8.60 & 9.90          \\
GS-212097 & 53.1628799 & -27.7672272 & 1.22  & 45.17               & 43.93 & Pa $\alpha$ & 4725 $\pm$ 65 & 8.34  &     10.83      \\
GS-196290 & 53.1488495 & -27.8211861 & 2.584 & 46.38               & 44.69 & He I & 1267 $\pm$ 69 & 7.58 &    11.10       \\ \hline    
GN-1000721     &       189.153763	  &  62.2223206       &    2.948    &   44.36     &  43.32 & He I & 3125 $\pm$ 1009 & 7.68 & 10.50\\
GN-1028801	  &       189.095581	  &  62.2574081       &    2.588     &   46.32     &  44.66 & Pa $\beta$ & 4941 $\pm$ 140 & 8.75 & 10.24\\
GN-1077827	  &       189.266602	  &  62.1992989       &    3.408  &   46.15     &  44.56 & He I & 4049 $\pm$ 1005 & 8.53 & 10.26\\
GN-1095877	  &       189.189438	  &  62.3136978       &    2.927 &   45.92     &  44.43 & He I & 3352 $\pm$ 303 & 8.30 & 11.26$^*$\\
GN-1025078	  &       189.152634	  &  62.22966         &    0.96    &   45.14     &  43.90 & Pa $\alpha$ & 2058 $\pm$ 54 & 7.61 & 9.98\\
GN-1095207	  &       189.278656	  &  62.2839203       &    1.022   &   45.63     &  44.25 & Pa $\alpha$ & 3460 $\pm$ 42 & 8.23 & 9.96$^*$\\
GN-1083261	  &       189.268066	  &  62.2461662       &    2.217    &   46.21     &  44.60 & He I & 1997 $\pm$ 342 & 7.93 & 10.18\\
GN-1094302	  &       189.346619	  &  62.2606583       &    2.244   &   45.31     &  44.03 & He I & 4884 $\pm$ 763 & 8.42 & 10.44$^*$\\
GN-1027287	  &       189.194733	  &  62.2460823       &    2.004   &   45.98     &  44.47 & He I & 5429 $\pm$ 1029 & 8.73 & 10.37\\
GN-1024921	  &       189.175369	  &  62.2253914       &    2.019  &   44.59     &  43.46 & Pa $\beta$ & 1453 $\pm$ 106 & 7.10 & 10.72\\
\enddata
\tablecomments{$^a$ ID is the combination of ``GS-" (GOODS-S) or ``GN-" (GOODS-N) with JADES ID; $^b$ AGN bolometric luminosity derived from SED fitting, except for GS-184451, 206907, and GN-1000721 whose $L_{bol}$ is converted from the observed X-ray 2-10 keV luminosity; $^c$ X-ray 2-10 keV luminosity converted from the $L_{bol}$ using the \citet{Duras2020} bolometric correction, except for GS-184451, 206907 and GN-1000721, which is the observed X-ray 2-10 keV luminosity from \citet{Luo2017}; $^d$ Intrinsic full-width half maximum (FWHM) of the broad component corrected for the instrumental and morphological broadening}; $^e$ Stellar mass derived by imaging decomposition and PSF-subtracted flux SED fitting, while those marked with ``$*$" are derived from SED decomposition due to lack of JWST/NIRCam wideband photometry.
\label{tab:sample}
\end{deluxetable*}

\section{Black hole and galaxy masses}
\label{cnproperties}

\subsection{Black Hole Masses} \label{sec:BHmass}
Masses of the black holes in our z $\sim 3$ AGNs were estimated by the Single-Epoch virial mass method, based on the velocity width of the broad line component and the AGN luminosity in a specific band \citep{Vestergaard2002}. \citet{Ricci2017} have compared the widths of the three lines we use (Pa $\alpha$, Pa $\beta$, and He\,I $\lambda$10833\,\AA) and find that for 90\% of their sample, He\,I widths track closely those in H $\alpha$; as expected, Pa $\alpha$ and Pa $\beta$ also closely track H $\alpha$. We therefore use the measured line widths interchangeably with no corrections. 


\begin{figure*}
\centering
    \includegraphics[width=1\textwidth]{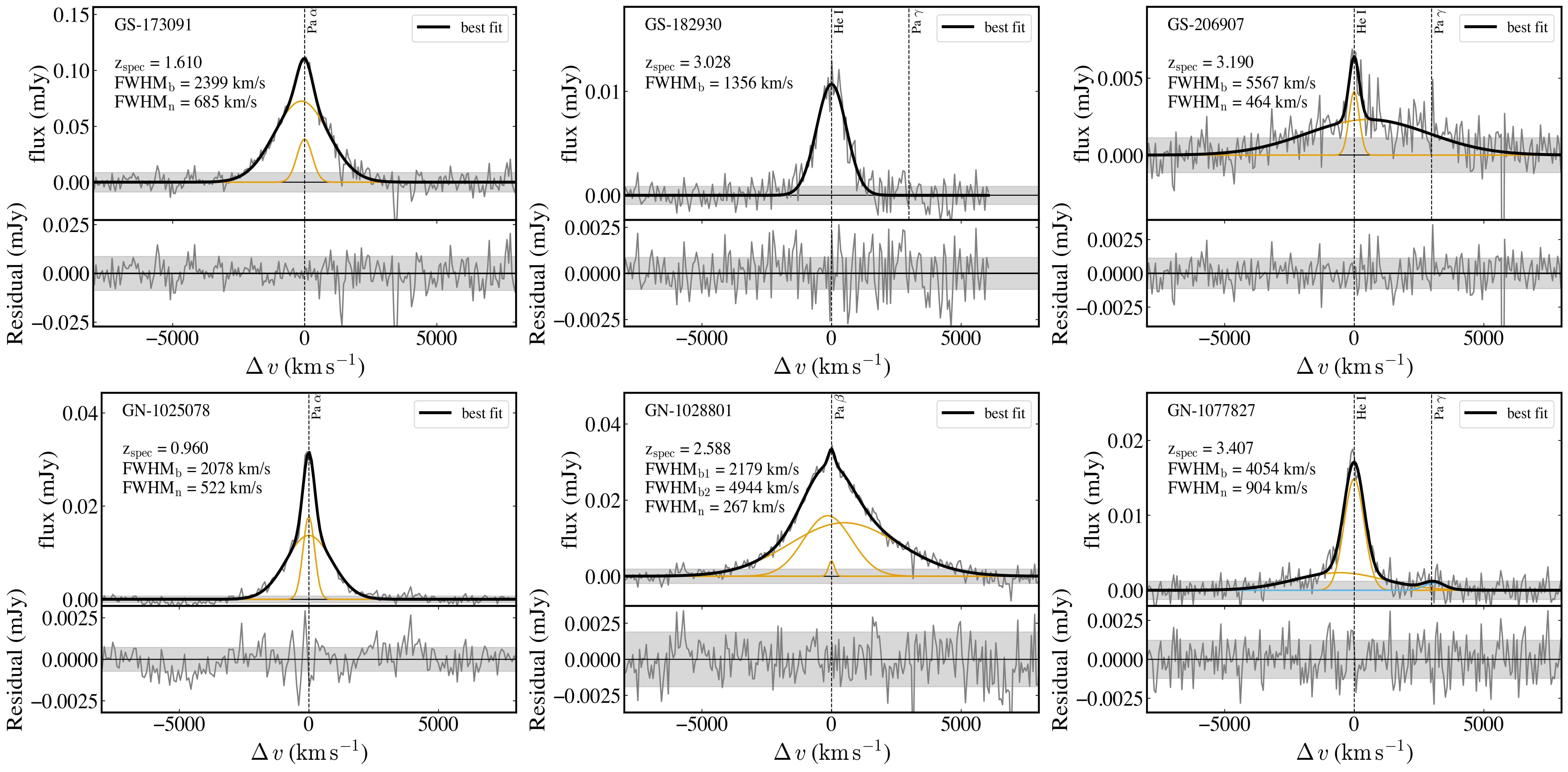}
\caption{Examples of line profile fitting for the broad Pa $\alpha$, Pa $\beta$, and He\,I $\lambda$10833\,\AA~lines. In each panel, we show the extracted 1D spectrum (continuum subtracted) and 1$\sigma$ error level (grey line and shading), the best-fit line profile (black) and each component (orange lines are for the targeted lines (Pa $\alpha$/Pa $\beta$/He I), while the blue line in the bottom right panel is an additional component for close Pa $\gamma$), and the residual of the best-fit.}
\label{Fig:linefitting_example}
\end{figure*}

We fitted the spectra to determine the central wavelength and the line widths. We began by using the pre-existing spectroscopic or photometric redshift to select the wavelength range containing the line ($\pm$5000 $\mathrm{km\,s^{-1}}$). Then, we set the peak wavelength as an initial guess for the central wavelength. We used one or two Gaussian components to model the profile of the broad NIR lines (Pa $\alpha$, Pa $\beta$, and He\,I $\lambda$10833\,\AA). 
Since the typical FWHM of the broad component is around 2000 $\mathrm{km\,s^{-1}}$ \citep[e.g.,][]{Landt2008}, we selected 0.05 $\micron$ wide regions to either side to define the continuum, lying beyond two to three times the FWHM of the broad component ($\pm 4000$-$6000$ $\mathrm{km\,s^{-1}}$, varied case by case to capture a clean continuum). We applied linear fits to these regions to determine the continuum level\footnote{For GS-196290, we only fit the red continuum region with a width of 0.1 $\micron$, since the targeted line lies at the blue end of the spectrum}. 

After subtracting the continuum, we used the Python package \texttt{lmfit} \citep{Newville2014} to fit the line profile. We first tried two Gaussian components. The center of the narrow component was allowed to shift within $\pm 200$ $\mathrm{km\,s^{-1}}$, while the center of the broad component was more flexible ($\pm 1000$ $\mathrm{km\,s^{-1}}$), given that previous works have found the broad component can be blue or red shifted relative to the narrow component \citep[e.g.,][]{Zastrocky2024}. We also required the FWHM of the narrow (broad) component to be no larger (smaller) than $1000$ $\mathrm{km\,s^{-1}}$ (except for GN-1000721 and 1095207 whose narrow component slightly exceeds this threshold). After visual inspection, we found the above procedure works for most of our AGNs, except: 
\begin{itemize}
    \item For GS-182930, 196290, and GN-1024921, we found their narrow component is either too faint compared to the broad component or too narrow ($< 50$ $\mathrm{km\,s^{-1}}$). Therefore, we think the signal of the narrow component of the line is washed out by the noise. Thus, we only used one Gaussian component to fit the line and treated it as the broad component.
    \item For GN-1027287, 1077827, 1083261, 1094302, and 1095877, the Paschen $\gamma$ line (10941\AA) has non-negligible emission and is blended with the targeted He\,I $\lambda$10833\,\AA~ line. Therefore, an additional Gaussian component was assigned for modelling Paschen $\gamma$, where the components have the same offset and width as the narrow component of the He\,I $\lambda$10833\,\AA\,line.
    \item For GN-1028801, an intermediate component needed to be assigned for the Paschen $\beta$ line to improve the fitting. Multiple broad components have been commonly seen in BL AGNs\citep[e.g.,][]{Ricci2022, Kuhn2024}, which might trace the complicated geometry and dynamics of the BLR of AGN \citep{Peterson2006,Popovic2019}.
\end{itemize}

Examples of the fitting results are shown in Figure~\ref{Fig:linefitting_example} (the rest of them are shown in the Appendix and Figure~\ref{Fig:spec_appendix}). The observed FWHMs of the broad components of our AGNs are between $1300$ and $6000$ $\mathrm{km\,s^{-1}}$. The spectroscopic redshift was also calculated based on the wavelength offset of the center of the narrow component (for two/three-component cases) or the broad component (for one-component cases) relative to its rest-frame wavelength.

Then, to quantify the measurement errors of the line profile parameters and the spectroscopic redshift, we applied a Monte Carlo (MC) simulation, i.e., generating 100 mock spectra based on the observed flux errors and re-fitting them to get the distributions of best-fit parameters.

Next, we derived the intrinsic FWHM of the broad component by correcting the instrumental and morphological broadening effects. Given the NIRCam/WFSS instrumental resolution at the wavelength range of 3--5 $\micron$ is about 1500, the typical instrumental broadening of FWHM ($\text{FWHM}_\text{inst}$) is $200~\mathrm{km\,s^{-1}}$. To evaluate the morphological broadening effect, we collapsed the imaging data of each source at the observed wavelength of the broad line along the dispersion direction, and derived a Gaussian convolution kernel such that the convolved PSF profile would match the observed brightness profile of the source. The angular FWHM of this convolution kernel then corresponds to an FWHM of the grism spectrum, which is the morphological broadening width of the observed broad spectral line ($\text{FWHM}_\text{morph}$). The FWHM of the convolution kernel for our samples ranges from 0.09--0.3\arcsec, corresponding to the $\text{FWHM}_\text{morph}$ range of 90--350 $\mathrm{km\,s^{-1}}$. We then derived the intrinsic FWHM of the broad component by $\text{FWHM}_\text{int}=\sqrt{\text{FWHM}_\text{obs}^2-\text{FWHM}_\text{inst}^2-\text{FWHM}_\text{morph}^2}$. We notice that such correction for our sample is small ($<50~\mathrm{km\,s^{-1}}$).

Then, using the errors derived from this MC simulation and then requiring the intrinsic FWHM of the broad component to be at least 1$\sigma$ higher than $1000~\mathrm{km\,s^{-1}}$ and the peak flux of the broad component to be 1$\sigma$ higher than the 1$\sigma$ error level of the spectrum, we confirmed all our 18 BL AGNs have a detected broad NIR line.

We finally used the virial mass estimation relation reported by \citet{Ricci2017}:
\begin{equation}
\begin{split}
    \log \left(\frac{M_{\bullet}}{M_{\odot}}\right)&=8.03+2 \log \left(\frac{\rm FWHM_{NIR, int}}{10^4 \mathrm{~km} \mathrm{~s}^{-1}}\right)\\
    &+0.5 \log \left(\frac{L_{\mathrm{X}}}{10^{42}\,\mathrm{~erg} \mathrm{~s}^{-1}}\right),
\end{split}
\end{equation}
where $L_{\mathrm{X}}$ is the AGN X-ray luminosity at 2-10 keV ($L_{\mathrm{2-10\,keV}}$). This relation was measured with a scatter of 0.4 dex and a virial factor of $f=4.31$. Also, this virial BH mass relation can be applied to all of the three NIR lines (Pa$\alpha$, Pa$\beta$ and He I) given that \citet{Ricci2017} found a good agreement between the FWHMs of H$\alpha$ and of those three NIR emission lines. $L_{\mathrm{X}}$ of our AGNs was determined as described at the end of Section~\ref{sec:AGNselect}.
The final $M_{\bullet}$ of our AGN sample ranges from $10^7$ to $10^9$ $M_{\odot}$, and the individual measurements are shown in Table~\ref{Tab1}. The typical uncertainty of the BH mass propagated from the uncertainty of the FWHM measurements is about 0.1 dex. Combined with the systematic uncertainty of \citet{Ricci2017}'s relation (0.4 dex), the overall uncertainty of our BH mass measurements is $\sim$0.4 dex.

In addition, for cases with more than one component, we can also measure the velocity offset of the broad component relative to the narrow one. We found, besides the known outflow host GS-197911, that there are six more BL AGNs (GS-184451, 173091, GN-1027285, 1095207, 1077827, and 1000721) with a broad component blueshifted more than $100$ $\mathrm{km\,s^{-1}}$ from the systemic velocity, which could be alternatively explained by galactic outflows rather than AGN BLRs. However, we rule out the possibility of outflow-dominated broadening for all of them:
\begin{itemize}
    \item For GS-184451, GN-1024921, 1027285, 1077827, and 1000721, we detected a nearby narrow forbidden line (e.g.,  [S\,III] $\lambda$5931\,\AA  or [Fe\,II] $\lambda$1.257, 1.644\,$\micron$) in the same grism spectrum, for which the width is much narrower than the broad component of the targeted BLR-indicator line. This indicates that their broad NIR lines are dominated by AGN BLR emission rather than outflows.
    \item For GS-173091 and GN-1095207, the grism spectra do not show any other significant emission lines except for the targeted NIR broad line. However, given that their broad components are only blueshifted by $\sim$100 and 400 $\mathrm{km\,s^{-1}}$, respectively, which are common in BL AGN observations \citep{Shen2016,Zastrocky2024}, and are dramatically broad (2399 and 3470 $\mathrm{km\,s^{-1}}$, respectively), we conclude that their broad lines are dominated by AGN BLR  as well.
\end{itemize}

\subsection{Host Galaxies } \label{sec:stellarmass}
In this section, we introduce the two approaches applied to measuring the stellar mass of the BL AGN hosts. For the AGNs that have JADES JWST/NIRCam broadband images, we conduct AGN-galaxy imaging decomposition to measure the fluxes of the hosts, then use SED fitting with a custom setup for the AGN component \citep{Lyu2022} in \texttt{Prospector} \citep{Johnson2021} to derive the stellar masses. For the four BL AGNs outside of the JADES footprint for which we cannot do the same imaging decomposition as the other 14 AGNs, we estimate the stellar masses by decomposing their UV to mid-IR SEDs with a modified \texttt{Prospector} SED fitting using the semi-empirical AGN component.

\subsubsection{AGN-Galaxy Decomposition} \label{sec:AGNdecomp}

\begin{figure*}
\centering
    \includegraphics[width=0.8\textwidth]{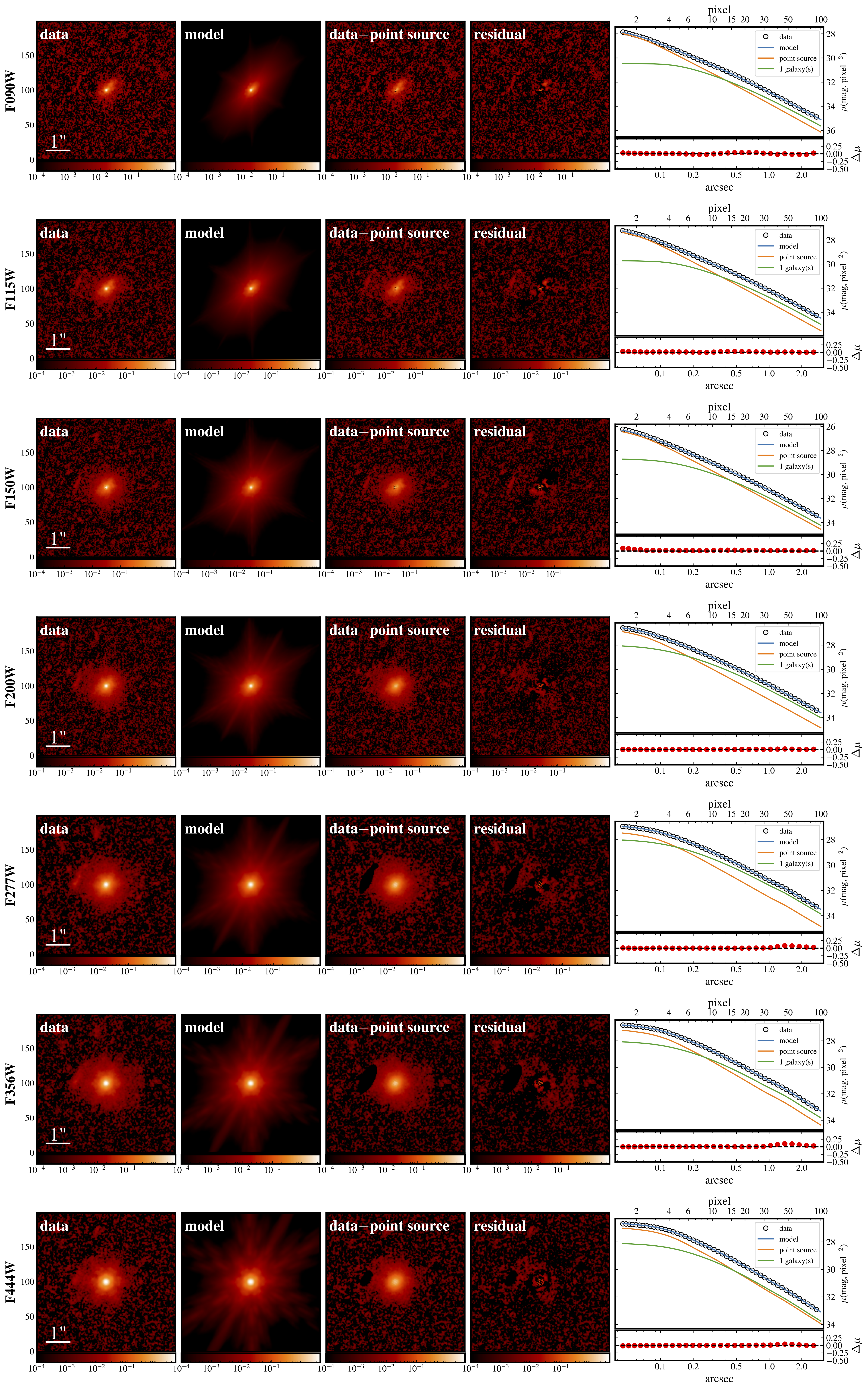}
\caption{Example \texttt{galight} fits of all 7 JWST/NIRCam broadband images (F090W to F444W from top to bottom) for GN-1083261. The columns from left to right show the data, model (PSF+S\'ersic), data $-$ point source (host galaxy), residual (data $-$ model) images, and 1D surface brightness profiles. From the data $-$ point source image, we confirmed the detection of the host galaxy.}
\label{Fig:galight_example}
\end{figure*}

We used the Python package \texttt{galight} \citep{Ding2020} to decompose the light from AGN and galaxy on JADES NIRCam broad-band images (from F090W to F444W), by fitting them with a PSF component and a S\'ersic component (with a free S\'ersic index (n)) convolved with a PSF. We used a single S\'ersic profile; although this approach is not able to determine the substructure if any, our primary goal is to obtain the overall mass of the galaxy, which is not sensitive to the detailed morphology.

First, the target images were cut out from all of the JADES JWST/NIRCam broadband drizzled images in the GOODS-S and GOODS-N field (with a nominal resolution of $\sim$0.03''), with a FOV of 6\arcsec\ $\times$ 6\arcsec\ to make sure all light coming from our sources is included. The PSFs applied during the fitting are from the JADES collaboration (see Appendix A in \citet{Ji2023} for details). All nearby objects (detected in the segmentation map at $S/N>3$ in the JADES photometric catalog) were masked before fitting. There is a bright companion to the southeast of GS-206907 shown in all NIRCam images, even though it can only be deblended by the segmentation maps from F090W to F277W. Therefore, we used an additional S\'ersic model to fit it simultaneously in these five short wavelength bands, instead of masking.

The full fitting routine supported by \texttt{galight} starts with the Particle Swarm Optimizer (PSO; \citealt{Kennedy1995}) to find the best-fit model. Then, the inferred minimized parameters are passed to the Markov Chain Monte Carlo (MCMC) routine to estimate the posterior parameter distributions. The values at the peak of each posterior parameter distribution are the final best-fit parameters. Finally, the fluxes of the AGNs and host galaxies are measured by summing up the model light of the best-fit PSF and S\'ersic components, respectively.


Figure~\ref{Fig:galight_example} shows an example of the \texttt{galight} fits in the seven NIRCam broadband images. The ``data - point source" images reveal the light from the host galaxies, and the 1D surface brightness profiles tell us whether the AGN or the galaxy dominates the total brightnesses of the sources. By visually inspecting the host galaxy images (``data - point source") and then requiring the host to be well-detected at least in the rest-frame $1\mu m$ band, we confirmed that
the hosts of the 14 BL AGNs that have images covering all NIRCam broadbands are well-detected. In addition, the ``residual" images in Figure~\ref{Fig:galight_example} illustrate the difference between the \texttt{galight} best-fit model (PSF + S\'ersic) and the observed fluxes. 

We also tested the robustness of our galaxy flux measurements by subtracting the \texttt{galight} best-fit point source flux from the PSF-convolved KRON fluxes for the galaxies from the JADES DR2 catalog\footnote{\url{https://archive.stsci.edu/hlsp/jades}}. We confirmed that this measure of galaxy flux (JADES total flux - PSF flux) is similar to the S\'ersic galaxy flux. 

The MCMC results provide the uncertainties of the best-fit parameters, but there are some concerns that they might be underestimated \citep{Ding2023, Tanaka2024}. Therefore, we used the MC method to derive the errors to see if they agree with the MCMC errors reported by \texttt{galight}. We generated 100 mock images for each target based on the observed flux errors and re-fitted them to get the distributions of best-fit parameters. Overall, the MC error of the galaxy magnitude is comparable to the MCMC one, which is typically $\sim$ 0.01 mag. Also, the error in stellar mass propagated from the galaxy flux error is $<$0.1 dex, which is much smaller than the error associated with the SED fitting to convert the flux to mass ($\sim$0.3 dex, see Section~\ref{sec:sedfitting}). Therefore, we still use the MCMC errors as the galaxy flux errors. 

Even though studying host galaxy properties besides stellar mass, e.g., morphology and size, is beyond the scope of this work, we briefly point out here that the distribution of the S\'ersic indices, especially those in the long wavelength filters, peaks at n $\sim$ 1--2. 
However, as \citet{Krywult2017} have pointed out, the S\'ersic index for bulge and disk galaxies decreases with redshift. Namely, n $\sim$ 1 at $1<z<4$ does not necessarily mean the galaxies are disk-like. Therefore, we are conservative about whether our AGNs are mostly hosted by late-type (disk-dominated) galaxies, although the host galaxy type (i.e. evolutionary stage) of the AGN sample at different redshifts, is critical for studying the time evolution of $M_\bullet$-$M_*$ relation (see Section~\ref{sec:discuss}).
Meanwhile, the best-fit galaxy effective radius of our 18 BL AGNs is typically 0.2\arcsec corresponding to physical scales of $\sim$1.5 kpc for galaxies at z$\sim$1--3.5. In this case, it is really difficult to tell the existence of a bulge or a disk, and further to separate its light contribution to the total galaxy flux to derive the bulge or disk mass. Therefore, we are only able to study the $M_\bullet$-$M_*$ relation for our BL AGN sample at $1<z<4$, rather than the $M_\bullet$-$M_{b}$ relation.  

\subsubsection{Galaxy Stellar Mass Measured by Prospector SED Fitting}
\label{sec:sedfitting}

\begin{figure}
\centering
    \includegraphics[width=1\columnwidth]{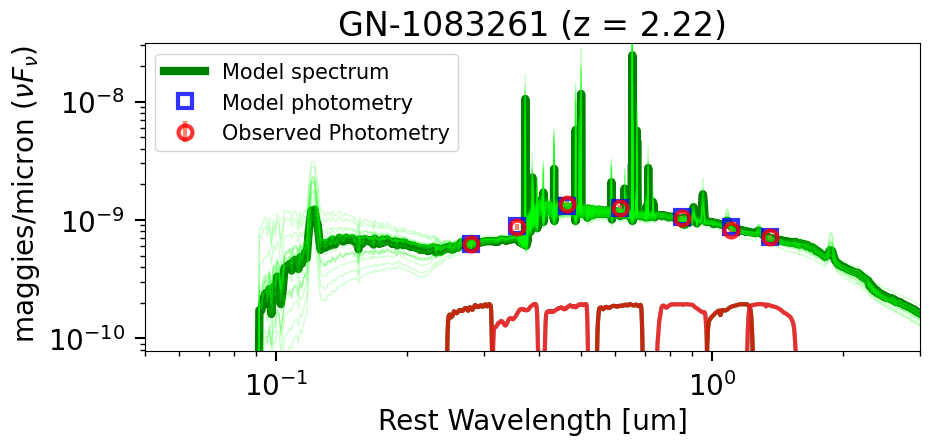}
\caption{Example of the \texttt{Prospector} galaxy SED fitting for GN-1083261, using the PSF-subtracted fluxes derived from \texttt{galight}. The PSF-subtracted fluxes (red circles) are fitted with a SED model (green lines) that contains the stellar component and the nebular emissions. The blue squares represent the best-fit model photometry. The bottom red enclosed regions show the transmission curves of F090W, F115W, F150W, F200W, F277W, F356W, and F444W from left to right.}
\label{Fig:sed_example}
\end{figure}

The stellar masses of the AGN hosts were measured by fitting the PSF-subtracted fluxes with the software \texttt{Prospector}. A delayed-tau star formation history with Kroupa initial mass function and the \citet{Kriek2013} extinction law for the stellar continuum were assumed. Since the point-source flux is removed by PSF subtraction, we did not add any AGN component, but only incorporated stellar components and associated nebular emission lines components and then derived the stellar mass from the best-fit models. One example of the galaxy SED fitting is shown in Figure~\ref{Fig:sed_example}. The typical statistical uncertainty of the \texttt{Prospector} stellar mass is less than 0.1-0.15 dex, which is smaller than the systematic errors introduced by different SED model assumptions such as star formation history (0.2-0.3 dex). As a result, we only consider systematic errors during the following analysis.

For GS-243640, GN-1095877, 1095207, and 1094302, which do not have sufficient multi-band NIRCam images for accurate imaging decomposition, we derived their stellar masses using SED decomposition, i.e, fitting the UV--to--mid-IR integrated galaxy emission separated from the total SED. We used the measurements based on CANDELS/SHARDS photometry from \citet{Barro2019} and fitted with the modified \texttt{Prospector} code described in \citet{Lyu2024}, where a semi-empirical model for AGN UV to mid-IR continua with nebular emission lines and dust attenuation is introduced to replace the default AGN torus model in \texttt{Prospector}. We confirmed that these four BL AGNs with stellar mass derived from a different method do not have any impacts on our results of the $M_\bullet$-$M_*$ relation. The stellar masses of our AGNs are listed in Table~\ref{Tab1}; overall they are hosted by massive galaxies ($\log(M_{*}/M_{\odot})\gtrsim10$).


\section{Results} \label{sec:results}

\subsection{Approach}
\label{foundations}

\begin{figure*}
\centering
    \includegraphics[width=0.8\textwidth]{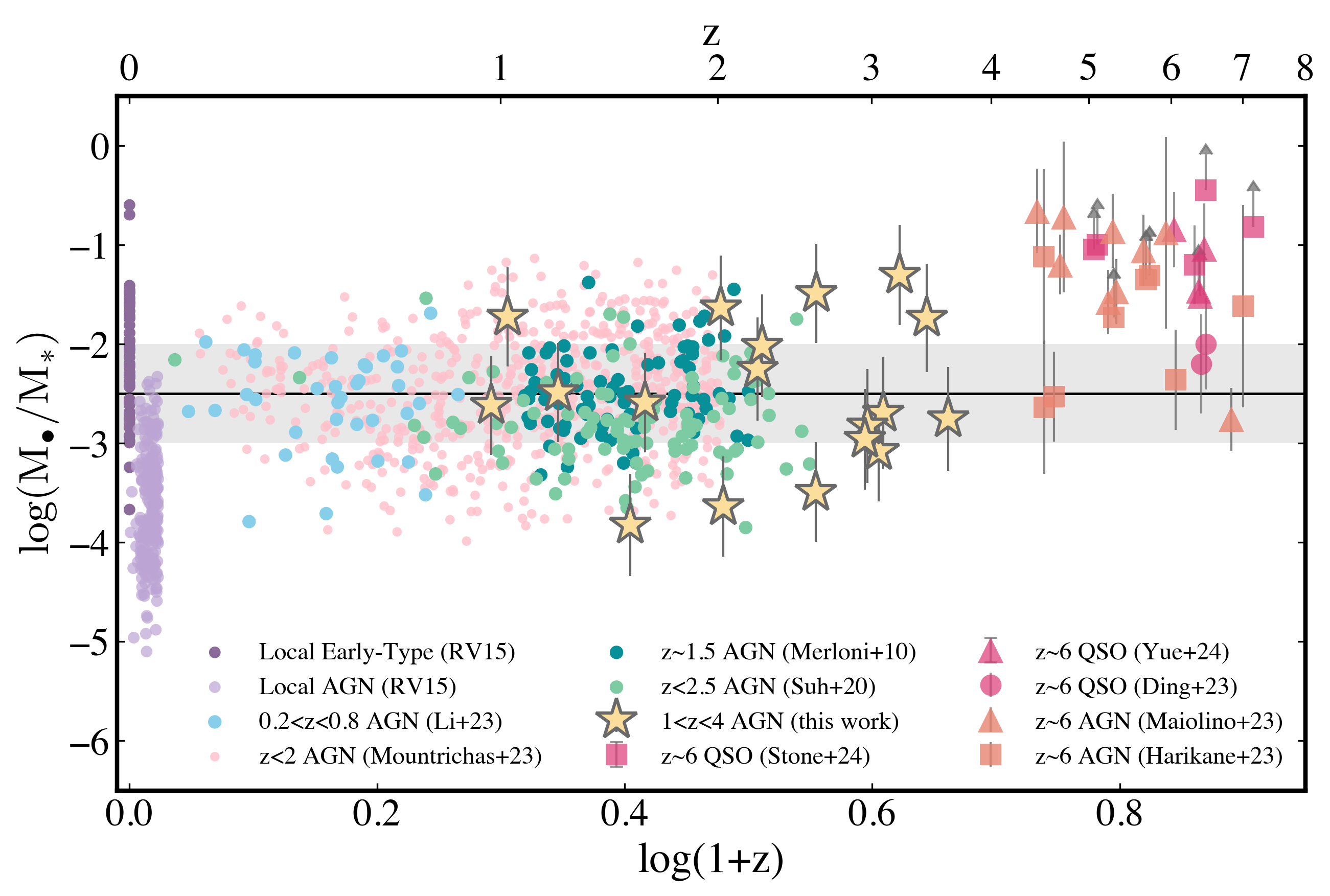}
\caption{Summary of the $M_\bullet/M_{*}$ behavior from the local Universe to $z=7$. 
We include the \citet{Merloni2010} and \citet{Suh2020} samples, along with local AGNs and early-type galaxies from \citet{Reines2015} (light and dark purple), $0.2<z<0.8$ AGNs (light blue) from \citet{JLi2023}, and $z<2$ AGNs (pink) from \citet{Mountrichas2023}. The $z>4$ faint AGNs (magenta) from \citet{Maiolino2023,Harikane2023} and quasars (orange) from \citet{Ding2023,Yue2024,Stone2024} are also shown as a reference. Our new BL AGN sample (yellow stars) at $1<z<4$ fills the redshift gap in mass scaling relation studies. The $z\sim1$ relation ($\log (M_\bullet/M_{*})=-2.5$ with a scatter of 0.5 dex) is represented by the black solid line and the gray shaded region. 
For $z \gtrsim 0.1$, the average $\log (M_\bullet/M_{*})$ values of massive AGNs fall well above the typical behavior for local AGNs (usually hosted by late-type galaxies) and are slightly below the behavior of local early-type galaxies \citep{Reines2015}. However, there is little change relative to this ratio for normal AGNs at $0.1 < z < 4$. Dwarf AGNs can have relatively higher ratios than massive AGNs at similar redshifts.}
\label{Fig:timeline}
\end{figure*}

To study the evolution of the $M_\bullet$-$M_{*}$ relation up to $z\sim4$,  we need to put our analysis on the same basic foundation as the studies for $z = 1$ to $2.5$. The $M_\bullet/M_{*}$ ratio is best behaved for the galaxy bulges \citep{Kormendy2013}, but for high redshift samples, it is not always possible to isolate galaxy bulge, especially beyond cosmic noon, as we pointed out in Section~\ref{sec:AGNdecomp}. In this case, relations to the integrated stellar output are used, particularly those due to \citet{Reines2015, Greene2020} developed for local galaxies. Since local AGNs are typically in late-type (disk-dominated, relatively lower-mass) galaxies, the relation for them is usually assumed. However, high redshift studies identify AGNs primarily in relatively massive, early-type (bulge-dominated) host galaxies. For example, most of the AGN hosts (26 out of 38) in \citet{JLi2023} at $z\sim$0.2--0.8 are early-type systems (light blue points in Figure~\ref{Fig:timeline}). Therefore, the studies of AGN hosts at z $\gtrsim 0.2$ find that $M_\bullet/M_{*}$ for them lies well above the local late-type/AGN relation from \citet{Reines2015}, approaching the relation for early-type galaxies (see Figure~\ref{Fig:timeline}). To maintain consistency, studies of the BH-galaxy mass relation should be based on a reasonable reference $M_\bullet/M_{*}$ ratio appropriate for the inspected stellar mass range. In section~\ref{sec:discuss}, we will discuss the issue of the $M_\bullet/M_{*}$ ratio comparison for samples at different evolutionary stages. 

A second consideration is how to quantify the observational biases \citep[e.g.,][]{Lauer2007}. We do this with a MC program similar to that introduced by \citet{Li2021} in which the observations are simulated according to (1) a distribution of galaxy masses (i.e. stellar mass function (SMF)); (2) an assumed intrinsic behavior of $M_\bullet/M_{*}$ with a scatter; (3) a BL AGN fraction (fraction of galaxies containing a BL AGN); (4) an input Eddington Ratio Distribution Function (ERDF); (5) measurement uncertainties for stellar mass and black hole mass.
Each MC trial is for a mock galaxy having a true stellar mass from the SMF but with observables according to an intrinsic $M_\bullet/M_{*}$ ratio sampled from the assumed intrinsic relation, BL AGN fraction, ERDF, and errors. The trial then yields a mock galaxy with observed stellar mass and black hole mass, as well as AGN luminosity and broad line width if it hosts an active nucleus. By applying the observational limit, we will obtain the ensemble of results distributed according to these uncertainties and subject to the ``Lauer bias''. 

A feature of our approach is that we simulate the actual observations we could take in an area-limited survey like FRESCO, i.e., with the appropriate number of BL AGNs for a single set of observations. Using this actual observation design, we compare the observed distribution of $\log (M_\bullet/M_*)$ with the simulated ones. 
We use the Kolmogorov-Smirnov test to determine the probability that the observed distribution could be drawn by chance from the distribution defined by the many simulated observations. 

\subsection{Determination of the $M_{\bullet}-M_{*}$ relation}
\label{sec:baseline}





The offset in the $M_{\bullet} - M_{*}$ relation in the local Universe relative to that {\it observed} in AGNs at moderately high redshift (Figure~\ref{Fig:timeline}) is likely due to a selection bias, but must be taken into account when evaluating evidence for evolution.
The different choices of the reference benchmark can cause disagreements in the existence of redshift evolution in $M_{\bullet} - M_{*}$ relation for z $\le$ 2, even though the derived intrinsic $M_{\bullet}/M_{*}$ ratios at $1<z<2$  generally agree with each other reasonably well. For example, for $1 < z < 2.5$, \citet{Suh2020} find that average ratio of log($M_\bullet/M_{*}$) is -2.64 for galaxies with $8.5 < \log(M_\bullet/M_{*})$ $< 9.5$ and -3.00 for galaxies with $7 < \log(M_\bullet/M_{*})$$ < 8.5$, and -2.50 when their sample is combined with the \citet{Merloni2010} sample. Similarly, without differentiating for black hole mass, \citet{Mountrichas2023} finds an intrinsic average value of log($M_\bullet/M_{*}$) $\sim$ -2.50 and \citet{Setoguchi2021} obtain -2.22 for this redshift range. These values are consistent with each other, while not comparable with the ratio often used from \citet{Reines2015} for which $\log(M_\bullet/M_{*})\sim -3.83$. 
We notice that a dozen rare dwarf AGN-host galaxies ($\log(M_{*}/M_{\odot})\lesssim9.5$) recently reported by \citet{Mezcua2023,Mezcua2024} have significantly higher $M_\bullet/M_{*}$ ratios ($\log(M_\bullet/M_{*})\sim -1.5$) compared to the values for massive AGN hosts. Given that this work focuses on the mass scaling relation at the massive end, we do not include those dwarf BL AGNs in the comparison. We will briefly discuss their different $M_\bullet/M_{*}$ behaviors in Section~\ref{sec:discuss}.

Therefore, to capture the range of observed values for massive galaxies for $1 < z < 2.5$, we will carry out analyses for log($M_\bullet/M_{*}$) = -2.50 with an intrinsic scatter of 0.5 dex as the baseline (hereafter we refer to this as the ``$z\sim1$ relation"), and also with log($M_\bullet/M_{*}$) = -3.00. These two values are appropriate for comparison with the other high redshift studies at a similar stellar mass range, and using both gives assurance that our conclusions are not strongly dependent on the choice of fiducial mass ratio.

The average BH-to-galaxy mass ratio ($\log (M_{\bullet}$/$M_{*})$) for our BL AGN sample, ($\log(M_\bullet/M_{*})=-2.61^{+0.90}_{-0.59}$), is consistent with the ratio for the z $\sim$ 1 relation even before correcting for the observational biases (see the bias test in Section~\ref{sec:obsbias}).

\subsection{Modeling observational biases}
\label{sec:obsbias}

In this section, we apply a Monte Carlo simulation to explore how observational biases affect the apparent $M_{\bullet} - M_{*}$ relation from our observations. 
Basically, we generate a mock AGN sample that represents the underlying AGN population at a specific redshift epoch and that follows the z $\sim$ 1 relation; we then apply the observation effects to this population to make a mock ``observable" AGN sample, and then we test how likely it is that the observed distribution on the $M_{\bullet} - M_{*}$ diagram is consistent with the mock distribution. 
In Section~\ref{sec:inf_survey}, we will first start with an ideal scenario in which the survey area is infinitely large to compare the intrinsic distribution and the observations on the $M_{\bullet} - M_{*}$ plane. In Section~\ref{sec:lim_survey}, we will simulate mock observations by restricting the simulated survey area to the size of our actual observations and applying a BL AGN fraction related to the whole galaxy population. The third and final step is to properly test how likely the z $\sim$ 1 relation can reproduce the observed distribution. We discuss the procedures for generating mock AGN populations in the Appendix.

\subsubsection{Infinitely large survey}
\label{sec:inf_survey}

\begin{figure*}
\centering
    \includegraphics[width=1\textwidth]{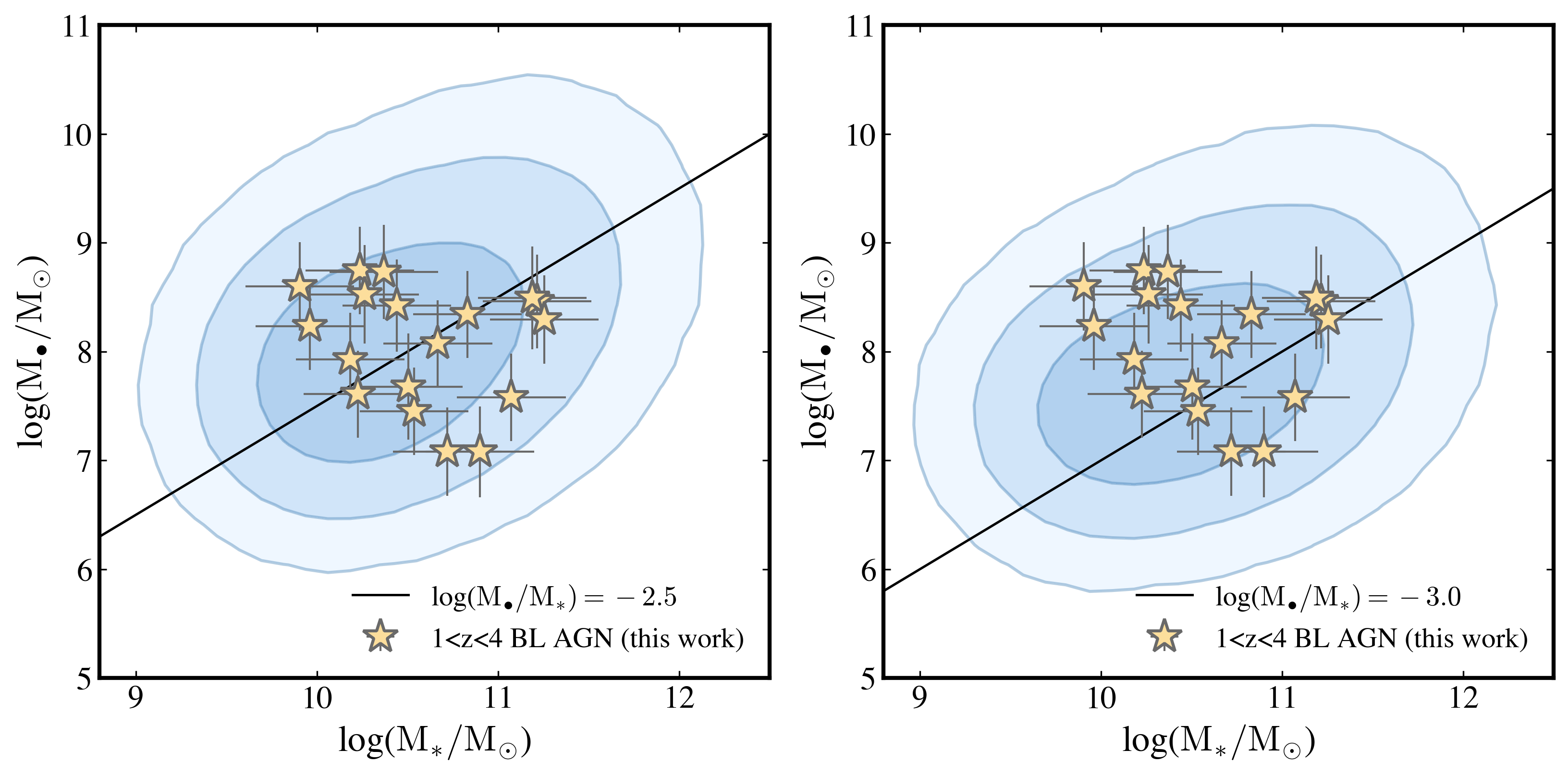}
\caption{The distributions on the $M_{\bullet}$-$M_{*}$ diagram at $1<z<4$  based on the assumption of $\log(M_{\bullet}/M_*)=-2.5$ (left) and $-3.0$ (right). The contours show the predicted distribution from our Monte Carlo simulation that includes both errors and biases. Specifically, the contours fall slightly above the input log($M_\bullet/M_{*}$) primarily because of the ``Lauer bias''.}
\label{Fig:observable_z2}
\end{figure*}



\begin{figure*}
\centering
    \includegraphics[width=0.8\textwidth]{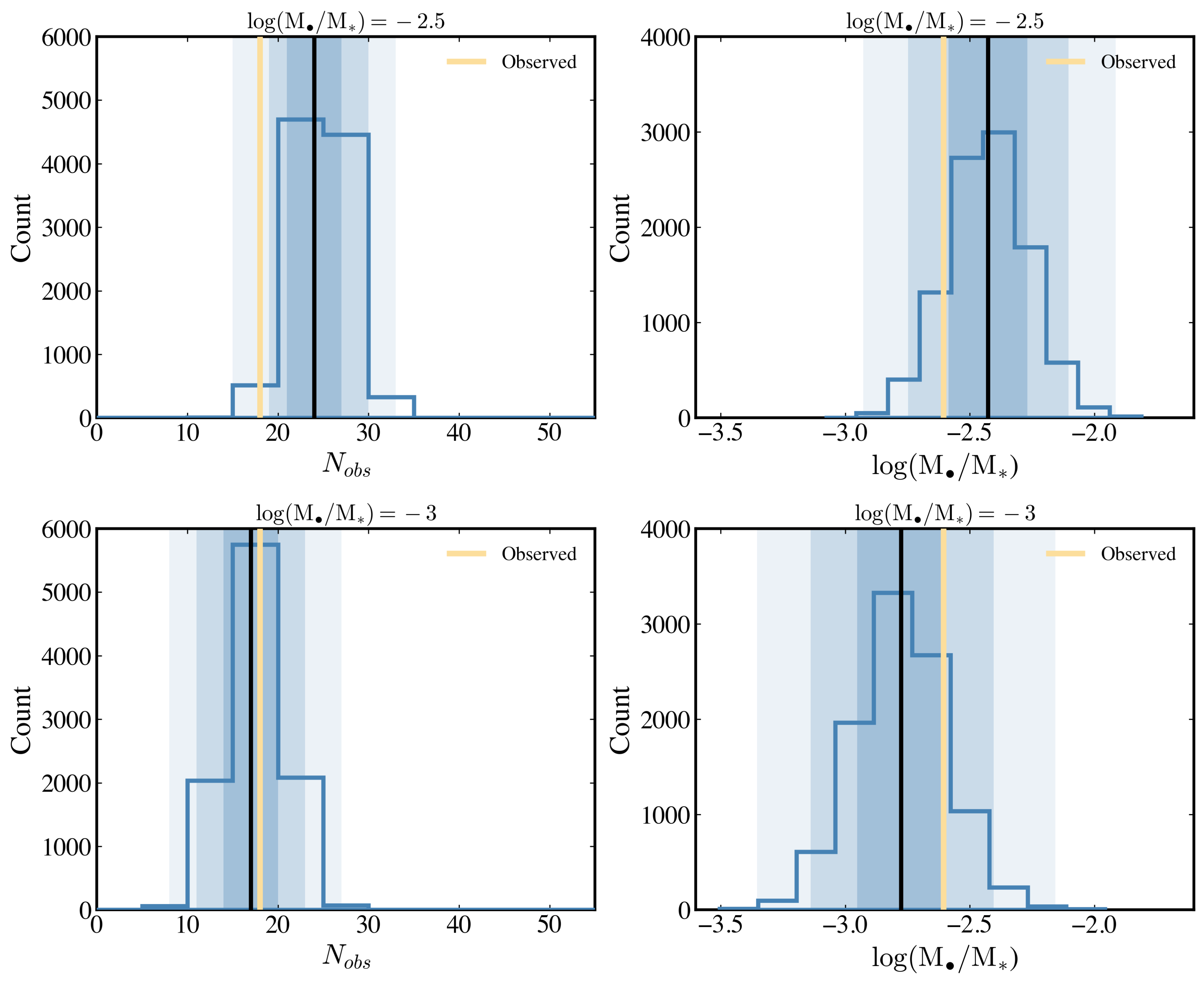}
\caption{MC simulation results for the mock AGN sample assuming the underlying intrinsic relation is either $\log (M_{\bullet}$/$M_{*})=-2.5$ (top) or $-3.0$ (bottom). The blue histograms represent the mock distributions, the black lines represent the median values of the mock distributions, and the yellow lines represent the observed value of our AGN sample.
{\bf Left:} The distribution of the simulated ``observable" mock AGN number count in a FRESCO-like area-limited survey. The observed number count ($N_{obs} = 18$) is slightly lower than the predicted number count if $\log (M_{\bullet}$/$M_{*})=-2.5$ ($N_{obs}=24$, the difference is $< 3\sigma$), while consistent with the predicted number count if $\log (M_{\bullet}$/$M_{*})=-3.0$ ($N_{obs}=17$, the difference is $<1\sigma$).
{\bf Right:} The average $\log (M_{\bullet}$/$M_{*})$ distribution (blue open histograms) for 10000 mock observations in the FRESCO fields, for which the ranges of $M_{*}$ and $z$ are matched with our observed BL AGN sample. The black lines represent the median $\log (M_{\bullet}$/$M_{*})$ of the observed AGNs. The difference between the peak values of the distribution for the simulated ``observable" mock AGNs and the observed $\log (M_{\bullet}$/$M_{*})$ ratio of our $1<z<4$ AGN sample has a significance of less than 2$\sigma$ and 1$\sigma$, respectively, for the two simulations. 
Overall, the results show that our observed BL AGNs at $1<z<4$ are consistent with the simulated ``observable" sample with either $\log (M_{\bullet}$/$M_{*})=-2.5$ or $-3.0$. }
\label{Fig:MC_all}
\end{figure*}


\begin{figure}
\centering
    \includegraphics[width=1\columnwidth]{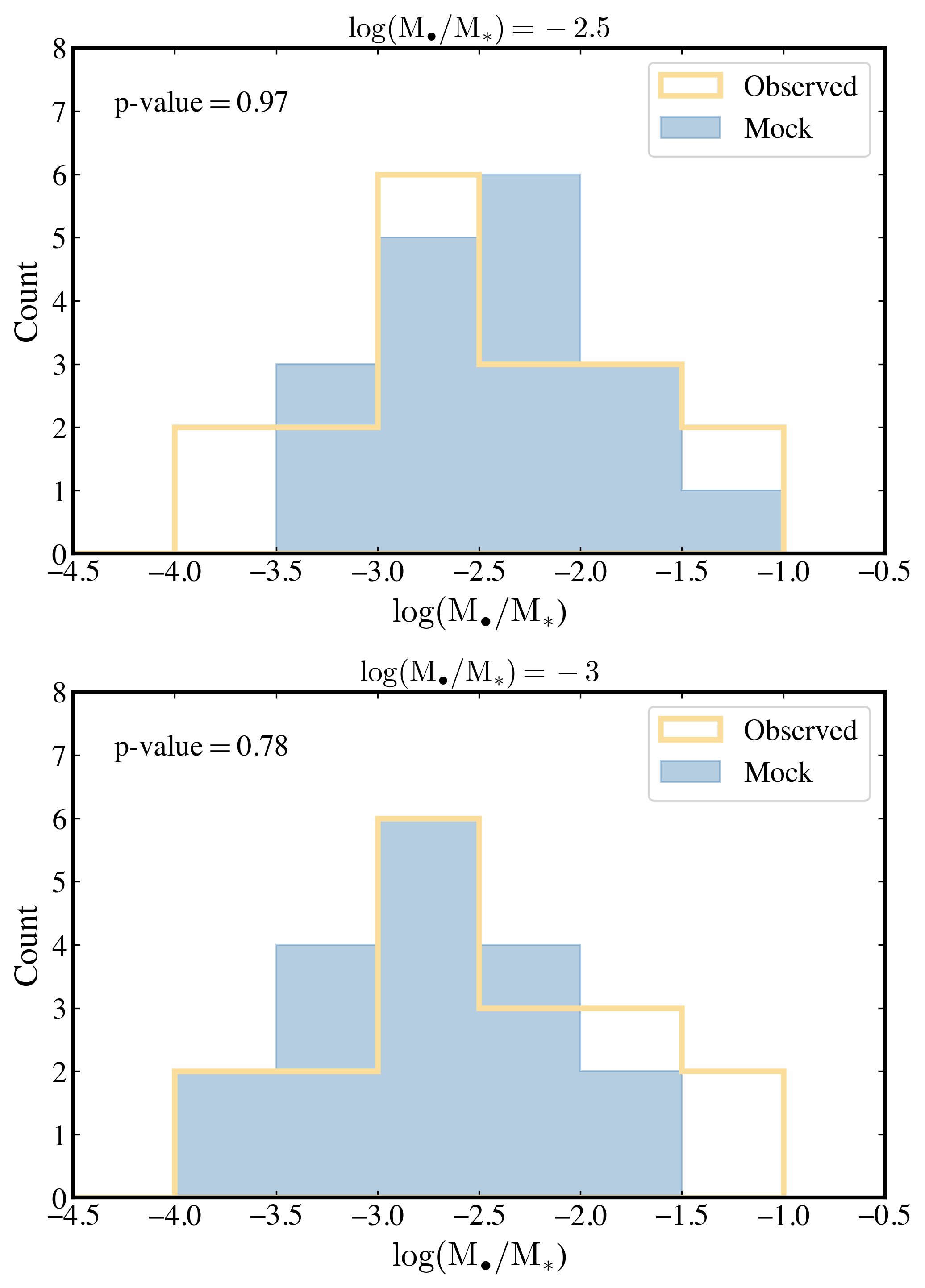}
\caption{Comparison between the reference distribution of $\log (M_{\bullet}$/$M_{*})$ derived by averaging all the mock observations (blue-filled histogram) and the observed distribution (yellow-open histogram), assuming the underlying intrinsic relation is either $\log (M_{\bullet}$/$M_{*})=-2.5$ (top) or $-3.0$ (bottom). The p-value of the K-S test suggests the two distributions have no significant difference, supporting the idea of no-evolution of the $M_{\bullet}$/$M_{*}$ ratio from z$=1$ to $=$4.
}
\label{Fig:KStest_average_all}
\end{figure}


We use $10^7$ trials for randomly creating $10^7$ mock BL AGNs with true stellar mass $M_{*\text{, true}}$ and redshift $z_{\text{true}}$ from the galaxy stellar mass function (SMF, \citealt{Weaver2023}). This arbitrary number is much larger than the BL AGN population size in the FRESCO fields ($\sim10^2$, see Section~\ref{sec:lim_survey}), so it can represent an infinitely large survey scenario.
Assuming that the BL AGNs follow the $z \sim 1$ relation, we determine the true BH mass $M_{\bullet, true}$ and then the bolometric luminosity ($L_{bol}$) by adopting the intrinsic Eddington ratio distribution function (ERDF) at $z \sim 2.15$ by \citet{KS2013}.

Next, assuming the virial BH mass and stellar mass measurements are not biased from the true mass but are subject only to measurement errors, we randomly add a Gaussian error with a dispersion of 0.4 dex to the $M_{\bullet\text{, true}}$, and 0.3 dex to the $M_{*\text{, true}}$ to obtain $M_{\bullet\text{, obs}}$ and $M_{*\text{, obs}}$. These errors were determined from our mass measurement errors mentioned in Section~\ref{sec:BHmass} and \ref{sec:stellarmass}.

Finally, we apply the observational limits to select the ``observable" mock AGNs.
The first observational bias comes from the BL AGN selection. In this work, only the AGNs that have a measurable broad line (FWHM$>1000\,\mathrm{km\,s^{-1}}$) can be identified as broad-line AGNs and be used for single-epoch $M_{\bullet}$ estimation. To estimate the FWHM of the broad line for mock AGNs, we use $M_{\bullet\text{, true}}$ and $L_{bol}$ based on the \citet{Ricci2017} single-epoch BH mass relation. Another limit arises because of the spectral sensitivity of FRESCO/CONGRESS. Typically, they can reach to a line flux sensitivity of $\sim5\times10^{-18}\,\mathrm{erg\,s^{-1}\,cm^{-2}}$ for observing a broad line of 1000$\,\mathrm{km\,s^{-1}}$ (but shallower for detecting a broader line (see Appendix)). We convert the line flux sensitivity to the spectroscopic AGN bolometric luminosity limit $L_{\text{lim, spec}}$ based on $z_\text{true}$ (see Appendix for details). Therefore, the mock AGNs with $L_{bol}<L_{\text{lim, spec}}$ are undetectable. 
Considering the two observational limits mentioned above, we simulate an ``observable" BL AGN sample from the whole mock BL AGN sample.

The left panel of Figure~\ref{Fig:observable_z2} shows that the ``observable" $M_{\bullet\text{, obs}} - M_{*\text{, obs}}$ distribution at $1<z<4$ is only slightly biased upwards from the intrinsic z $\sim$ 1 relation ($\log (M_{\bullet}/M_{*})=-2.5$), indicating that our BL AGN sample at this redshift range is not much affected by observational biases. 
Also, all of our BL AGNs are enclosed within the $2\sigma$ contour of the ``observable" distribution; namely, our mock AGN simulation can successfully reproduce the observed AGNs without invoking any evolution in $M_\bullet/M_{*}$. 

To test the robustness of the no-evolution statement, we also check the result if we adopt $\log (M_{\bullet}/M_{*})=-3.0$ as the z $\sim$ 1 relation. We find, even for this lower ratio, our $1<z<4$ BL AGNs are still located in the central region of the ``observable" distribution.
This simulation and the results are very similar to the recent paper by \citet{Li2024}, which showed that with many trials (i.e., our ``infinite'' sample) the characteristics of $4 < z < 7$ AGNs recently claimed to be above the local $M_\bullet/M_{*}$ relation could be reproduced without requiring evolution in this ratio.

\subsubsection{Area-limited Survey}
\label{sec:lim_survey}

 The fact that the data points fall within a distribution does not necessarily indicate that these data points are consistent with the distribution. With a large enough sample, the distribution can successfully include the outliers, which may just reproduce some observed targets by chance. Moreover, in practice, our $1<z<4$ BL AGNs are selected from area-limited surveys rather than an infinitely large survey such as we modeled in Section~\ref{sec:inf_survey}: the $1<z<4$ BL AGNs are blindly selected from FRESCO and CONGRESS \footnote{The FRESCO survey mapped the 61 arcmin$^2$ areas in both the GOODS-S field and GOODS-N field, and the CONGRESS survey covers a very similar footprint as the FRESCO GOODS-N field.}.

Therefore, to further quantify the probability that the BL AGNs observed in the FRESCO and CONGRESS areas follow the mock distribution on the $M_{\bullet} - M_{*}$ plane assuming the z $\sim$ 1 relation, we ran another MC simulation to model the observations of the BL AGN population within a FRESCO-like survey area. In this case, we are doing an apples-to-apples comparison in which the mock AGN observations will have a comparable sample size and parameter ranges ($M_{*}$ and $z$) as the actual observations.

First, using the SMF, we estimate the expected number of $1<z<4$ (i.e. the redshift range of our BL AGN sample) galaxies appearing in the footprint of FRESCO's survey area (122 arcmin$^2$) to be $\sim1.3\times10^3$ over the stellar mass range of $\log {M_{*}/M_{\odot}}>9.9$.
Next, we run this number of trials to simulate the mock galaxy population following the z $\sim$ 1 relation ($\log (M_{\bullet}/M_{*})=-2.5$). 
We then apply a BL AGN fraction (the number ratio of BL AGNs to galaxies) from \citet{Schulze2015} ($3 ~\pm~ 1.2$\%, where the error is from their Figure 22) to this population to estimate the number of galaxies (or trials) to have a BL AGN ($\sim40 \pm 15$ BL AGNs).
After assigning an Eddington ratio, mass measurement uncertainties, and observational limits, as we did in Section~\ref{sec:inf_survey}, we generate a mock BL AGN observation. Finally, to exactly match the range of stellar mass between the mock observation and the true observation, we exclude the mock AGNs whose $M_{{*, obs}}$ is outside of the range of $9.9<\log(M_{*}/M_{\odot})<11.3$.

By repeating the procedure of mock observation generation 10000 times, the distribution of the ``observable'' AGN number counts in each mock observation with the intrinsic ratio of $\log (M_{\bullet}$/$M_{*})=-2.5$ is plotted in the top-left panel of Figure~\ref{Fig:MC_all}, which demonstrates that the median number count of the simulated ``observable" BL AGNs at z $\sim$ 2 within a FRESCO-like survey area is about $24 \pm 3$; the observed number count ($N_{obs}=18$) is within the expected $3\sigma$ error although slightly less than the predicted median. The generally successful matching of observable BL AGN number count between the simulation and observation confirms that our simulation accurately incorporates the correct assumptions of properties of the underlying galaxy and AGN populations (i.e., SMF, ERDF, BL AGN fraction), as well as the observational biases and measurement uncertainties. A marginally lower observed BL AGN number count compared to the prediction is not surprising given that we did not inspect every spectrum in FRESCO and CONGRESS, and we might also have rejected a few AGNs in our line width test.

Next, we test how likely our observed $1<z<4$ BL AGNs are statistically consistent with the mock observations following the $z \sim 1$ relation ($\log (M_{\bullet}$/$M_{*})=-2.5$) by comparing their $M_{\bullet}/M_{*}$ distributions. The top-right panel of Figure ~\ref{Fig:MC_all} illustrates the distribution of the median $\log (M_{\bullet}$/$M_{*})$ of mock observations; the observed median $\log (M_{\bullet}/M_{*})$ falls within the 2$\sigma$ range of the distribution of the mock observations. The bottom panels show the same information for $\log (M_{\bullet}$/$M_{*})=-3.0$; again there is no evidence for evolution. 

\subsection{Kolmogorov-Smirnov tests}

To further compare the full $M_{\bullet}/M_{*}$ distribution rather than just the median, we apply a Kolmogorov-Smirnov (K-S) test between the mock observations and the actual one to investigate their consistency. We use two different approaches of the K-S test to make this comparison:
\begin{enumerate}
     \item {\bf One-Sample Test} For this test, we average all the mock observations to obtain a noise-free reference distribution and do a K-S test between this reference and the actual observations. If the p-value of the K-S test is higher than 0.05, we can reject the hypothesis that the observed AGNs are not consistent with the mock reference distribution. To make the reference distribution, we randomly select 18 AGNs from each mock observation if the predicted number count is no less than 18, which also more or less accounts for the potential incompleteness of our BL AGN sample. We rank the 18 $\log (M_{\bullet}/M_{*})$ values of all mock observations and then average all the mock values at each rank to get the final averaged distribution.\footnote{We discard the mock observations with a predicted number count of less than 18 (1\% for the $log(M_\bullet/M_{*}) = -2.5$ case and 50\% for the $log(M_\bullet/M_{*}) = -3.0$ case) because it can not be ranked with other mock observations including 18 AGNs.} This approach reduces the uncertainty of the reference distribution introduced by the noise of each random simulated observation, providing a virtually noiseless distribution of observations from 18 targets.
    \item {\bf Two-Sample Test} In this case, we do a K-S test between each mock observation and the actual observation, namely, we make 10000 K-S tests and calculate the fraction of test results for which we can reject the hypothesis that the observed AGN sample does not come from the same distribution as the mock AGNs (the p-value of the K-S test is higher than 0.05). This approach fully accounts for the fluctuations that exist in our small sample of observations and provides a statistical probability of rejecting the null hypothesis.
   
\end{enumerate}

The one-sample test returns a p-value of 0.97 for $\log (M_{\bullet}$/$M_{*})$ = -2.5 (see Figure~\ref{Fig:KStest_average_all}).
In $>99\%$ (9916/10000) of the realizations of the two-sample test for $\log (M_{\bullet}$/$M_{*})$ = -2.5, the p-values are $>$ 0.05, indicating we can reject the hypothesis that the observed $\log (M_{\bullet}$/$M_{*})$ distribution is different from the mock distribution. All tests strongly indicate that there is no statistically significant difference of the observed $1<z<4$ sample from the expected no-evolution behavior at z $\sim$ 1 -- 2.5. 

Also, we tested the case of $\log(M_\bullet/M_{*}) = -3.0$ to see if the large dispersion of the $M_\bullet/M_{*}$ ratio at z $\sim 1 - 2.5$ could result in a different conclusion on the $M_\bullet/M_{*}$ evolution. 
The MC distributions of mock BL AGN number count and median $\log(M_\bullet/M_{*})$ are shown in the bottom panels of Figure ~\ref{Fig:MC_all}, and the K-S test between the reference mock $\log(M_\bullet/M_{*})$ distribution and the observed one is plotted in the bottom panel of Figure~\ref{Fig:KStest_average_all}, with a p-value of 0.78. Also, $>96\%$ (9676/10000) of the realizations of the two-sample test have a p-value higher than 0.05.
We confirmed that, even for $\log(M_\bullet/M_{*})=-3.0$, there is no statistically significant deviation of the observed sample from the expected no-evolution behavior at z $\sim$ 1 -- 2.5. 

\begin{figure*}
\centering

    \includegraphics[width=0.7\textwidth]{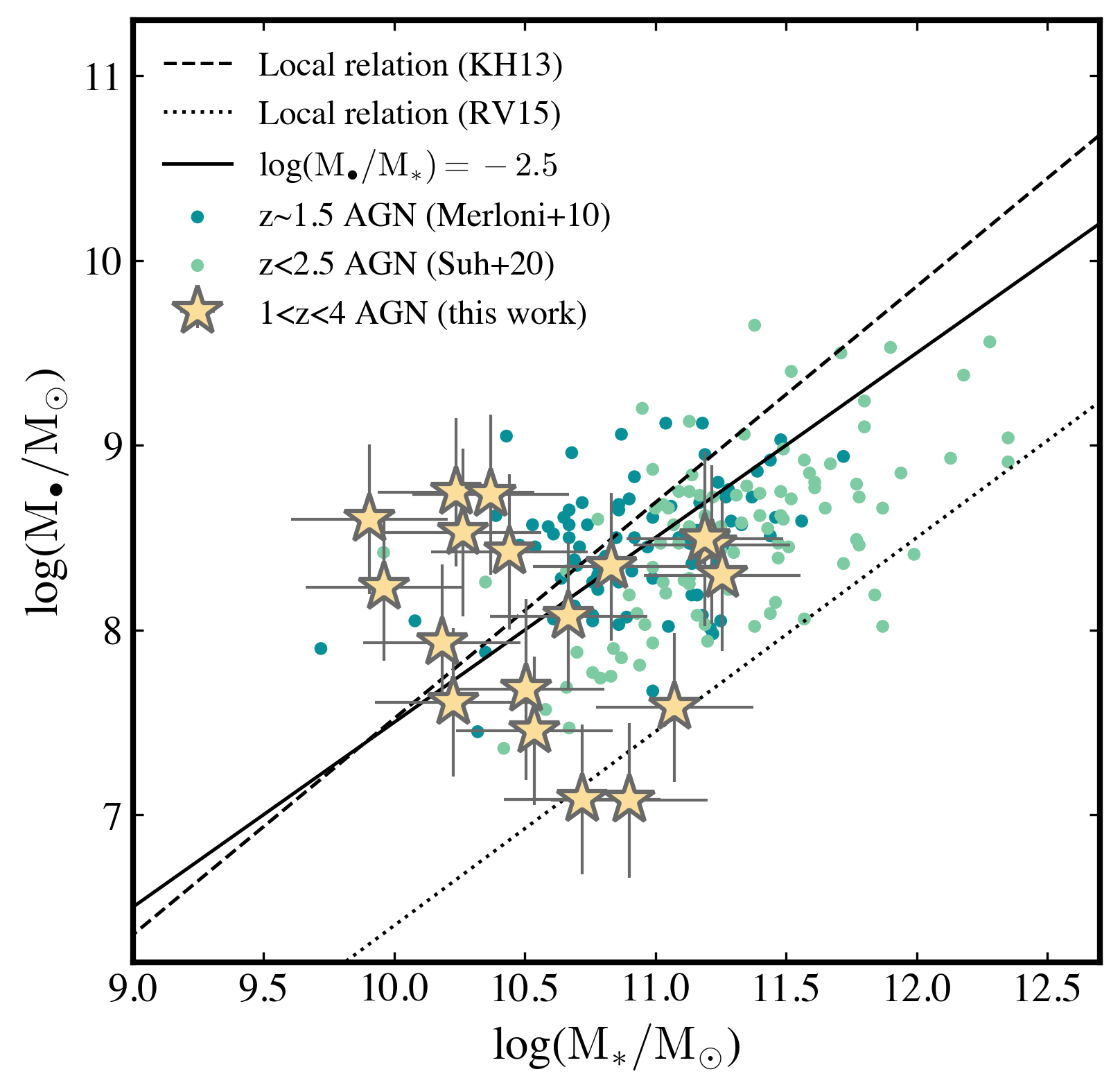}
\caption{$M_\bullet$ versus $M_{*}$ relation for the GOODS-S and GOODS-N AGN sample at $1<z<4$ (yellow stars). As a comparison, the AGN samples at cosmic noon ($z\sim 2$) from  \citet{Merloni2010} (dark green dots) and from \citet{Suh2020} (light green dots) are plotted. The local $M_{\bullet}-M_*$ relation for the bulge-dominated galaxies from \citet{Kormendy2013} and that for the AGN hosts from \citet{Reines2015} are illustrated by the black dashed line and the black dotted line, respectively. Also, we show the intrinsic BH-to-stellar mass ratio at $1<z<2$ ($\log(M_{\bullet}/M_*)=-2.5$) determined by previous studies (e.g., \citealt{Suh2020,Mountrichas2023}) using the solid black line.}
\label{Fig:massrelation}
\end{figure*}

Therefore, the observed $1<z<4$ BL AGNs are consistent with the mock distribution that is based on the z $\sim$ 1 relation after taking into account the selection biases, suggesting no evolution of the $M_\bullet/M_{*}$ ratio at the massive end ($\log(M_{*}/M_{\odot})>10$) at least up to z $\sim$ 4. We have already shown this result in Figure~\ref{Fig:timeline}. Figure~\ref{Fig:massrelation} is another presentation that highlights the lack of change in the relation between $M_\bullet$ and $M_{*}$ at $\log(M_{*}/M_{\odot})>10$ between the previously studied lower redshift AGNs and our sample.   

\section{Discussion} \label{sec:discuss}

Determining the masses of high redshift AGN host galaxies directly from their stellar populations or stellar dynamics, in a similar manner to the basis for the determination of the local $M_{\bullet}/M_{*}$ relation, is paramount for characterizing the possible evolution of the SMBH-galaxy correlations. As a step toward this goal, we have shown that the observed $M_{\bullet}/M_{*}$ ratio for massive galaxies has no significant change from z $\sim$ 1 to 4. Along with previous findings of no-evolution up to z$\sim$1 \citep[e.g.,][]{Suh2020,Mountrichas2023,JLi2023}, we conclude that the $M_{\bullet}-M_{*}$ relation at $\log(M_{*}/M_{\odot})>10$ up to z$\sim$4 is consistent with the local one.


Our study indicates some cautions on studying the $M_{\bullet}/M_{*}$ evolution using the local \citet{Reines2015} relation as the baseline. It should be noted that the local AGN sample in \citet{Reines2015} is based on lower-mass late-type systems, while AGNs at $1<z<4$ are usually detected in more massive galaxies. The higher-redshift AGN hosts may also be bulge-dominated galaxies  \citep[e.g.,][]{JLi2023}, although the morphology study of AGN hosts is still limited above z $\sim$ 1--2 even in the JWST era. Therefore, the shift in the mixture of galaxy types from the AGN sample in high-$z$ to that in the local Universe might be the reason for the significant change of the $M_\bullet/M_*$ ratio when comparing the value at high redshift with the local \citet{Reines2015} value. Future detailed discussion on this issue will be provided in a forthcoming paper.

In addition, some previous studies at $z < 2.5$ only used a $M_{\bullet}$/$M_{*}$ value to quantify the $M_{\bullet}$-$M_{*}$ relation, namely, simply assuming the $M_{\bullet}$/$M_{*}$ is independent of $M_\bullet$ or $M_*$ (the slope is always one).
Indeed, the relation from \citet{Reines2015}, 
\begin{equation}\label{eqn:mbh-mstar}
\log(M_\bullet/M_\odot) = \\
7.45 + 1.05 \times \log(M_{*}/10^{11}M_\odot),
\end{equation}
\noindent
suggests that $M_\bullet$ is almost linearly proportional to the stellar mass $M_*$, so a simple re-normalization might seem to be adequate. 
However, other work has found the $M_{\bullet}$/$M_{*}$ ratio can be different across a wide range of black hole masses or stellar masses. For example, \citet{Suh2020} found log($M_{\bullet}$/$M_{*}$) $\sim -2.64$ for $8.5 < \log(M_\bullet) < 9.5$ and log($M_{\bullet}$/$M_{*}$) $\sim-3.00$  for $7.0 < \log(M_\bullet) < 8.5$ at $z\sim1.5$. These observations indicate that not only the general host galaxy properties in that high-$z$ samples can be different from the low-$z$ samples (e.g., those studied in \citet{Reines2015}), but also that the range of black hole masses or stellar masses being sampled at high-$z$ are also different than the range at low-$z$ resulting in an intrinsic difference in the $M_\bullet/M_{*}$ ratios. Such behavior does not influence our results here since the stellar masses of our sample are relatively high and span a narrow range ($\sim$ 1 dex). However, it could complicate studies of low-mass BL AGNs ($\log(M_{*}/M_{\odot})\lesssim9.5$) beyond the local Universe, especially the new JWST discoveries of faint low-mass AGNs at z$>$4, where the host galaxy properties are even more challenging to determine.

In this work, we have also evaluated the net errors and biases in our measurements based on the Monte Carlo approach originally suggested by \citet{Li2021} with some additional improvements. We have shown that some considerably larger $M_{\bullet}$/$M_{*}$ ratios similar to those observed at high-$z$ can be produced by chance, sometimes after tens of millions of trials. However, the critical question is whether the observed ratios are likely in observational programs with a limited number of targets. The targets constitute the parent sample from which the observations are drawn, and a similar philosophy must be applied in the simulations. This distinction is important at high redshift where the available samples are small. 
In a forthcoming paper, we will extend the test of the redshift evolution of $M_{\bullet} - M_{*}$ relation up to z $\sim 7$ by using a similar analysis as developed in this paper. 

Lastly, as we pointed out in Section~\ref{sec:baseline}, this work only focuses on the $M_{\bullet}$-$M_{*}$ relation evolution between $1<z<4$ at the high-mass end because of the stellar mass range of our new BL AGN sample.  \citet{Mezcua2024} recently discovered a dozen BL AGNs hosted by dwarf star-forming galaxies at $1<z<3$. They found those low-mass AGN hosts are significantly above any of the local relations, which seems to indicate the presence of the $M_{\bullet}$-$M_{*}$ relation evolution at the dwarf regime. Also, when taking into account the future evolutionary pathway of those dwarf AGNs, the authors speculated that about 30\% of their dwarf AGNs could merge into the normal relation with an evolved and increased stellar mass at $z\sim0$. However, we do not include those dwarf AGN samples in our mass scaling relation evolution analysis and draw any conclusion at the low-mass regime in this paper, because 1) the local $M_{\bullet}$-$M_{*}$ baseline relation at the lower mass regime ($\log(M_{*}/M_{\odot})<9.5$) is still highly uncertain due to the limited number of dwarf BL AGNs observed with uniform BH mass measurements \citep{Reines2015,Greene2020}; 2) the higher redshift dwarf AGN sample size is also small and sparsely distributed across redshift (see Figure~\ref{Fig:timeline}); and  3) most importantly, those dwarf BL AGNs suffer from stronger observational biases compared to the massive ones (see Appendix~\ref{ap:mock_agn}), which were not evaluated in  \citet{Mezcua2023,Mezcua2024}. To better constrain the $M_{\bullet}$-$M_{*}$ relation at the low-mass end and examine its evolution with redshift, future observations of a larger sample of dwarf BL AGNs with consistent measurements of both BH mass and stellar mass, and a comprehensive understanding of their observational biases are essential.

\section{Conclusion} \label{sec:concl}

In this work, we built a new sample of 18 BL AGNs with broad rest-frame NIR spectral lines (Paschen lines and He\,I $\lambda$10833\,\AA) at $1<z<4$ in the GOODS-S and GOODS-N fields using the FRESCO and CONGRESS JWST/NIRCam grism spectra. We measure the BH masses of our BL AGN sample ($7.1<\log(M_{\bullet}/M_{\odot})<8.8$) using the single-epoch virial BH mass method based on the three NIR lines. We derived their stellar masses ($9.9<\log(M_{*}/M_{\odot})<11.3$) by either applying AGN-galaxy imaging decomposition on the JADES JWST/NIRCam broadband images and fitting the host-only SED, or doing the SED decomposition if targets are outside of the JADES footprints. 

We calculated the average $M_\bullet/M_{*}$ ratio for our new $1<z<4$ BL AGNs ($\log(M_\bullet/M_{*})=-2.61^{+0.90}_{-0.59}$) and found it is consistent with the values at z$\sim$1--2.5 from previous studies (Figure~\ref{Fig:timeline} and \ref{Fig:massrelation}). In other words, using the $z \sim 1$ relation ($\log(M_\bullet/M_{*}) = -2.5$) as the baseline, we do not observe an evolution of $M_\bullet/M_{*}$ ratio up to $z \sim 4$ (Figure~\ref{Fig:timeline}). We also tested the impact of observational biases on the observed $M_\bullet$-$M_{*}$ distribution by running an MC simulation to model both the mock $1<z<4$ AGN population in an infinitely large survey and the mock observations in a FRESCO-like (area-limited) survey, following the $z \sim 1$ relation. The observed $M_\bullet$-$M_{*}$ distribution is in good agreement with the intrinsic distribution derived with the infinite sample (Figure~\ref{Fig:observable_z2}).
Also, by applying the K-S test to compare the $\log(M_\bullet/M_{*})$ distribution between the mock observation and the true observation, we confirmed there is no significant difference between them (Figure~\ref{Fig:KStest_average_all}), which indicates that \textit{there is no evolution of $M_\bullet/M_{*}$ ratio at $\log(M_{*}/M_{\odot})\gtrsim10$ up to $z \sim 4$, even accounting for observational biases.}

This work sets up an appropriate baseline for comparison with the other high redshift studies and an improved MC tool to simulate mock observations with biases and errors for a fair comparison with true observations. In an accompanying paper, we will apply this developed tool to test the redshift evolution of $M_{\bullet} - M_{*}$ relation up to z $\sim 7$ by using a similar analysis as developed in this paper on higher-redshift ($4<z<7$) samples.

\section{Acknowledgments}
We thank Roberto Maiolino for helpful discussions and comments. This work was supported by the
JWST Mid-Infrared Instrument (MIRI) Science Team Lead 
grant, 80NSSC18K0555, from NASA Goddard Space Flight
Center to the University of Arizona. Y. S., Z. J., F. S., Y. Z., P. A. C., E. E., K. H., J. M. H., P. R., B. E. R., M. A. S., G. H. R., and C. N. A. W. also acknowledge support from the
NIRCam Science Team contract to the University of Arizona,
NAS 5-02015. I. S. acknowledges funding support from the Atracci\'on de Talento program, Grant No. 2022-T1/TIC-20472, of the Comunidad de
Madrid, Spain.  
S. T. acknowledges support by the Royal Society Research Grant G125142. The
work of C. C. W. is supported by NOIRLab, which is managed
by the Association of Universities for Research in Astronomy
(AURA) under a cooperative agreement with the National
Science Foundation. A. J. B. has received funding from the European
Research Council (ERC) under the European Union's Horizon
2020 Advanced Grant 789056 ``First Galaxies''.  F. D. E., and J. S.  acknowledge support by the
Science and Technology Facilities Council (STFC) and by the
ERC through Advanced Grant 695671 ``QUENCH", and the UKRI Frontier Research grant RISEandFALL.

This work is based on observations made with the
NASA/ESA/CSA James Webb Space Telescope. The
data were obtained from the Mikulski Archive for Space
Telescopes (MAST) at the Space Telescope Science Institute, which is operated by the Association of Universities for Research in Astronomy, Inc., under NASA
contract NAS 5-03127 for JWST. These observations are
associated with programs 1180, 1181, 1895, and 3577, for which the data can be accessed via \dataset[https://doi.org/10.17909/8tdj-8n28]{https://doi.org/10.17909/8tdj-8n28} \citep{https://doi.org/10.17909/8tdj-8n28}, and \dataset[https://doi.org/10.17909/gdyc-7g80]{https://doi.org/10.17909/gdyc-7g80} \citep{https://doi.org/10.17909/gdyc-7g80}. The authors sincerely thank the FRESCO team
(PI: Pascal Oesch) for
developing and executing their observing program.

\facility{JWST (NIRCam)}

\software{\texttt{AstroPy}\citep{astropy2013,astropy2018,astropy2022}, \texttt{galight}\citep{Ding2020}, \texttt{lmfit}\citep{Newville2014}, \texttt{Prospector}\citep{Johnson2021}, \texttt{SciPy}\citep{Virtanen2020}}

%

\appendix
\section{Broad-Line Fitting of the BL AGN}
\begin{figure*}[htb]
\figurenum{A1}
\centering
\includegraphics[width=1\textwidth]{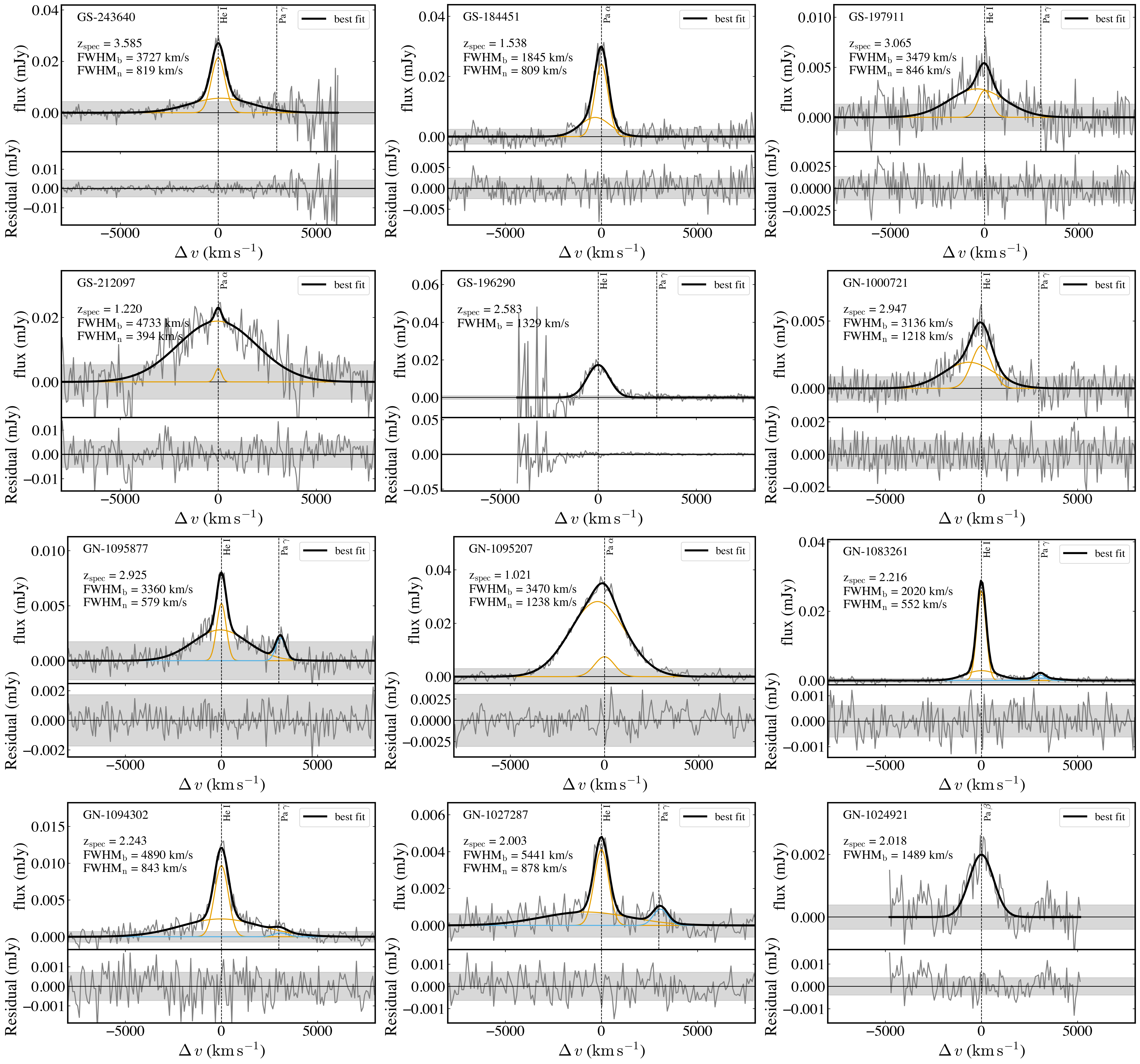}
\caption{Line profile fitting on the broad Pa $\alpha$, Pa $\beta$, and He\,I $\lambda$10833\,\AA~lines, for the rest (12/18) of our BL AGNs identified in this work, same as Figure~\ref{Fig:linefitting_example}.}
\label{Fig:spec_appendix}
\end{figure*}
We use the Python package \texttt{lmfit} \citep{Newville2014} to model the broad line (Pa $\alpha$, Pa $\beta$, and He\,I $\lambda$10833\,\AA~lines) profile of our BL AGNs. Besides the six examples shown in Figure~\ref{Fig:linefitting_example}, the rest (12) of the 18 BL AGNs are shown in Figure~\ref{Fig:spec_appendix}.

\section{Details of building a mock AGN population}
\label{ap:mock_agn}

\begin{figure*}
\figurenum{B1}
\centering
    \includegraphics[width=0.9\textwidth]{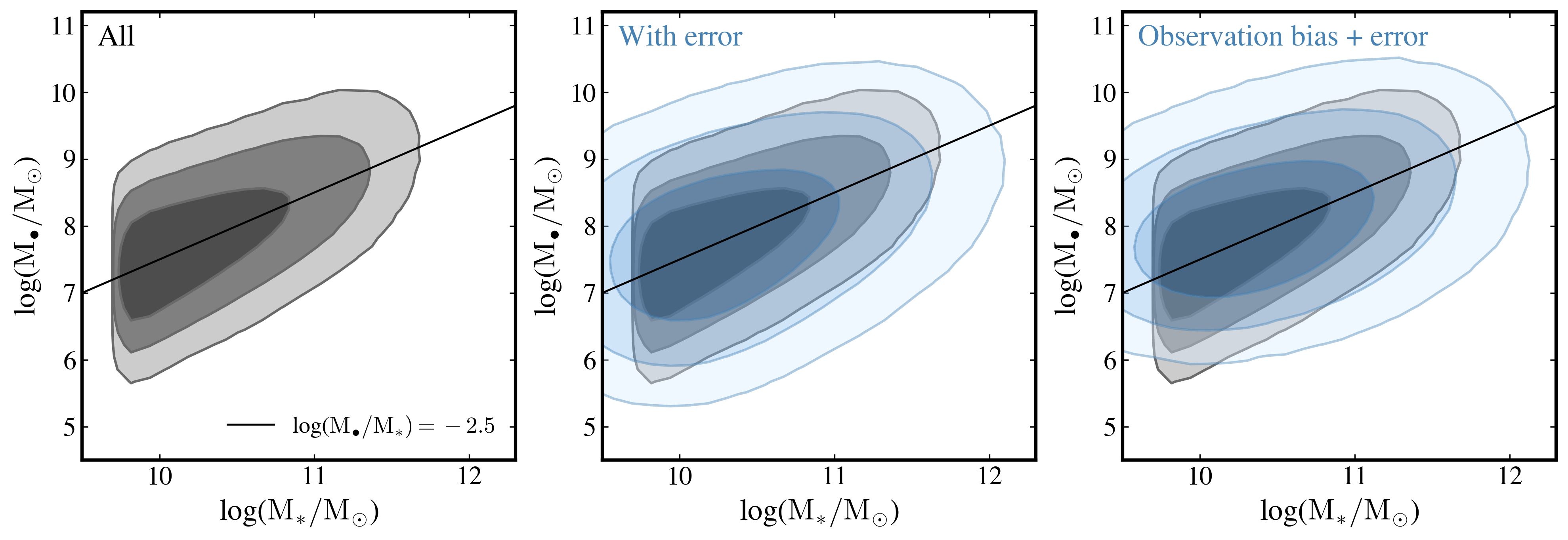}
\caption{The distributions of mock $1<z<4$ AGNs that follow the $z \sim 1$ relation on the $M_{\bullet}$-$M_{*}$ diagram. The grey contour in all panels represents the $M_{\bullet\text{, true}}$-$M_{*\text{, true}}$ distribution of the whole mock AGN population, without applying mass measurement errors. In the middle column, the blue contours show the $M_{\bullet\text{, obs}}$-$M_{*\text{, obs}}$ distributions of the whole mock AGN populations, namely, the observed distributions with the errors of $M_{\bullet}$ (0.4 dex),  and $M_{*}$ (0.3 dex). In the right panel, the blue contours show the $M_{\bullet\text{, obs}}$-$M_{*\text{, obs}}$ distributions of the ``observable" mock AGN populations, i.e., the errors of $M_{\bullet}$ and $M_{*}$, and the observational limits are added. All of the contours indicate the 1$\sigma$ to 3$\sigma$ levels.}
\label{Fig:selectioneffect}
\end{figure*}

For the infinitely large survey scenario (Section~\ref{sec:inf_survey}), to build a mock AGN population, we first randomly generate $10^7$ mock galaxies with true stellar mass $M_{*\text{, true}}$ and redshift $z_{\text{true}}$ from the COSMOS2020 stellar mass function (SMF) at $z
\sim 2$ \citep{Weaver2023}. The mock stellar mass is sampled from $10^{9.9}$ to $10^{12}\,M_{\odot}$ with the grid size of 0.005 dex in logarithmic scale, and the redshift $z_{\text{true}}$ is sampled in steps of $\Delta z=0.01$ within $1<z<4$ (corresponding to the redshift range of our AGN sample).
We limit the lower boundary of simulated stellar mass to $10^{9.9}\,M_{\odot}$ to match with the lowest stellar mass of our AGN sample, because there should be an observational limit determining the lowest stellar mass of galaxies that we can detect and can be well-measured through imaging-decomposition using JWST/NIRCam data. However, since it is hard to quantify and thus is not included in the observational limits we carefully model in this work, we simply limit the simulated stellar mass above a threshold to model this observational limit. 

In the next step, we assign a true BH mass $M_{\bullet\text{, true}}$ for each $M_{*\text{, true}}$ by assuming the $z \sim 1$ relation with the intrinsic scatter $\sigma_{\text{int}}$:
\begin{equation}
\begin{split}
&P\left(\log {M}_{\bullet}|\log {M}_{*}\right)=\frac{1}{\sqrt{2 \pi} \sigma_{\text{int}}} \\
&\times \exp \left(-\frac{\left(\log {M}_{\bullet}-\left(\log {M}_{*}+a\right)\right)^2}{2 \sigma_{\text{int}}^2}\right), 
\end{split}
\end{equation}
where $a$ is the intrinsic $M_\bullet/M_*$ ratio ($\log (M_\bullet/M_*)$).
Here we use the z $\sim$ 1 relation for AGNs, which have $a=-2.5$ or $-3.0$ and $\sigma_{\text{int}}=$ 0.5 dex.  
Then, we simply assume all of the generated mock galaxies have an AGN, given that we are modeling an infinitely large survey. In most cases, the AGNs are at very low luminosity and undetectable, as reflected in the ERDF. The true distribution of our mock AGN sample in the $M_{\bullet}$ versus $M_{*}$ plane is shown in the left panel of Figure~\ref{Fig:selectioneffect}. For each mock AGN, we then calculate its bolometric luminosity $L_{\text{bol, mock}}$ by assigning an Eddington ratio ($ \lambda_{\text{Edd}}=L_{\text{Edd}}/L_{\text{bol}}$) that is randomly sampled from the $z\sim 2.15$ ERDF by \citet{KS2013} and also 
the X-ray luminosity at 2-10 keV($L_{\text{2-10 keV}}$) using the X-ray bolometric correction reported by \citet{Duras2020}.

We further consider the uncertainty of the observed mass measurements ($M_{\bullet\text{, obs}}$ and $M_{*\text{, obs}}$) related to the true masses ($M_{\bullet\text{, true}}$ and $M_{*\text{, true}}$). In our case, the uncertainty of the virial BH mass method built by \citet{Ricci2017} is about 0.4 dex, and the SED-based stellar mass measurement error is about 0.3 dex. By adding random Gaussian errors to both $M_{\bullet\text{, true}}$ and $M_{*\text{, true}}$ we obtain $M_{\bullet\text{, obs}}$ and $M_{*\text{, obs}}$. The $M_{\bullet\text{, obs}}$-$M_{*\text{, obs}}$ distribution of the mock AGN sample is shown in the middle panel of Figure~\ref{Fig:selectioneffect}, which is significantly broader than the initial distribution.

Finally, we apply the observational limits to select the ``observable" mock AGNs. 
The first observational bias comes from the BL AGN selection. In this work, only the AGNs that have measurable broad line (FWHM$>1000\,\mathrm{km\,s^{-1}}$) can be identified as BL AGNs and be used for single-epoch $M_{\bullet}$ estimation. Therefore, we use $M_{\bullet\text{, mock}}$ and $L_{\text{2-10 keV}}$ to estimate the FWHM of the targeted NIR line based on the \citet{Ricci2017} single-epoch BH mass relation for each mock AGN. The mock AGNs with a broad line width less than $1000\,\mathrm{km\,s^{-1}}$ will be left out of the ``observable'' sample.
Another limit arises because of the spectral sensitivity of FRESCO. It is about $5\times10^{-18}\,\mathrm{erg\,s^{-1}\,cm^{-2}}$ for detecting a broad component with a width of $1000\,\mathrm{km\,s^{-1}}$. For detecting a broader line, the flux line sensitivity decreases as ${(\text{FWHM}/200)}^{0.5} \times2\times10^{-18}\,\mathrm{erg\,s^{-1}\,cm^{-2}}$, where $200\,\mathrm{km\,s^{-1}}$ is the instrumental broadening with spectral resolution $R \sim 1500$. We convert the line flux sensitivity to the line luminosity limit based on $z_{\text{true}}$. Given that most of our BL AGN samples are identified using the FRESCO F444W grism spectra with a wavelength coverage of 3.1--4.0 $\micron$, we assume the broad lines that we could detect in the FRESCO spectra for mock AGNs at $1<z<2$, $2<z<2.5$, and $2.5<z<4$ are Pa $\alpha$, Pa $\beta$, and He\,I $\lambda$10833\,\AA, respectively. By fitting the luminosity relation between H$\alpha$ and those three NIR lines using the \citet{Landt2008} local BL AGN sample, we can convert the targeted line luminosity limit to the H$\alpha$ line luminosity limit. We further convert the H$\alpha$ line luminosity to the spectroscopic AGN bolometric luminosity limit $L_{\text{lim, spec}}$ using the relation from \citet{Greene2005}. We classify the mock AGNs with $L_{\text{bol, obs}}<L_{\text{lim, spec}}$ as undetectable. Considering the two observational limits mentioned above, we made an ``observable" mock AGN subsample from the whole AGN sample. The final ``observable" $M_{\bullet\text{, obs}}$-$M_{*\text{, obs}}$ distribution is shown in the right panel of Figure~\ref{Fig:selectioneffect}. Comparing the ``observable" $M_{\bullet\text{, obs}}$-$M_{*\text{, obs}}$ distribution (blue) with the whole AGN $M_{\bullet\text{, true}}$-$M_{*\text{, true}}$ distribution (black) is straightforward to illustrate the significant effects of observational limits on biasing the real mass scaling relation to the observed one, which is known as the ``Lauer bias'' \citep{Lauer2007}.

For simulating the actual observation in a FRESCO-like survey (Section~\ref{sec:lim_survey}), we apply a similar routine to that for the infinitely large survey scenario. The only two differences are: (1) we now simulate the exact galaxy population within a FRESCO-like survey area instead of an arbitrarily large population. We first estimate the galaxy population size within a FRESCO-like survey field (122 arcmin$^2$) using the COSMOS2020 SMF, which results in 4170 galaxies with stellar mass ranging from $10^9$ to $10^{12}$$M_{\odot}$ and redshift in the range of $1<z<4$. (2) To estimate the BL AGN population underlying the whole galaxy population, we assume the BL AGN fraction at $1<z<4$ as a constant $ 3 \pm 1.2\%$ \citep{Schulze2015} of the whole galaxy sample, given that \citet{Schulze2015} found the BL AGN fraction at z $\sim$ 2 is almost constant when $\log ({M}_{\bullet}/M_{\odot})<9$.

\bibliography{reference}{}

\begin{thebibliography}{}
\expandafter\ifx\csname natexlab\endcsname\relax\def\natexlab#1{#1}\fi
\providecommand{\url}[1]{\href{#1}{#1}}
\providecommand{\dodoi}[1]{doi:~\href{http://doi.org/#1}{\nolinkurl{#1}}}
\providecommand{\doeprint}[1]{\href{http://ascl.net/#1}{\nolinkurl{http://ascl.net/#1}}}
\providecommand{\doarXiv}[1]{\href{https://arxiv.org/abs/#1}{\nolinkurl{https://arxiv.org/abs/#1}}}

\bibitem[{{Alberts} {et~al.}(2024){Alberts}, {Lyu}, {Shivaei}, {Rieke}, {Perez-Gonzalez}, {Bonventura}, {Zhu}, {Helton}, {Ji}, {Morrison}, {Robertson}, {Stone}, {Sun}, {Williams}, \& {Willmer}}]{Alberts2024}
{Alberts}, S., {Lyu}, J., {Shivaei}, I., {et~al.} 2024, arXiv e-prints, arXiv:2405.15972, \dodoi{10.48550/arXiv.2405.15972}

\bibitem[{{Alexander} \& {Hickox}(2012)}]{Alexander2012}
{Alexander}, D.~M., \& {Hickox}, R.~C. 2012, \nar, 56, 93, \dodoi{10.1016/j.newar.2011.11.003}

\bibitem[{{Astropy Collaboration} {et~al.}(2013){Astropy Collaboration}, {Robitaille}, {Tollerud}, {Greenfield}, {Droettboom}, {Bray}, {Aldcroft}, {Davis}, {Ginsburg}, {Price-Whelan}, {Kerzendorf}, {Conley}, {Crighton}, {Barbary}, {Muna}, {Ferguson}, {Grollier}, {Parikh}, {Nair}, {Unther}, {Deil}, {Woillez}, {Conseil}, {Kramer}, {Turner}, {Singer}, {Fox}, {Weaver}, {Zabalza}, {Edwards}, {Azalee Bostroem}, {Burke}, {Casey}, {Crawford}, {Dencheva}, {Ely}, {Jenness}, {Labrie}, {Lim}, {Pierfederici}, {Pontzen}, {Ptak}, {Refsdal}, {Servillat}, \& {Streicher}}]{astropy2013}
{Astropy Collaboration}, {Robitaille}, T.~P., {Tollerud}, E.~J., {et~al.} 2013, \aap, 558, A33, \dodoi{10.1051/0004-6361/201322068}

\bibitem[{{Astropy Collaboration} {et~al.}(2018){Astropy Collaboration}, {Price-Whelan}, {Sip{\H{o}}cz}, {G{\"u}nther}, {Lim}, {Crawford}, {Conseil}, {Shupe}, {Craig}, {Dencheva}, {Ginsburg}, {VanderPlas}, {Bradley}, {P{\'e}rez-Su{\'a}rez}, {de Val-Borro}, {Aldcroft}, {Cruz}, {Robitaille}, {Tollerud}, {Ardelean}, {Babej}, {Bach}, {Bachetti}, {Bakanov}, {Bamford}, {Barentsen}, {Barmby}, {Baumbach}, {Berry}, {Biscani}, {Boquien}, {Bostroem}, {Bouma}, {Brammer}, {Bray}, {Breytenbach}, {Buddelmeijer}, {Burke}, {Calderone}, {Cano Rodr{\'\i}guez}, {Cara}, {Cardoso}, {Cheedella}, {Copin}, {Corrales}, {Crichton}, {D'Avella}, {Deil}, {Depagne}, {Dietrich}, {Donath}, {Droettboom}, {Earl}, {Erben}, {Fabbro}, {Ferreira}, {Finethy}, {Fox}, {Garrison}, {Gibbons}, {Goldstein}, {Gommers}, {Greco}, {Greenfield}, {Groener}, {Grollier}, {Hagen}, {Hirst}, {Homeier}, {Horton}, {Hosseinzadeh}, {Hu}, {Hunkeler}, {Ivezi{\'c}}, {Jain}, {Jenness}, {Kanarek}, {Kendrew}, {Kern}, {Kerzendorf}, {Khvalko}, {King}, {Kirkby}, {Kulkarni},
  {Kumar}, {Lee}, {Lenz}, {Littlefair}, {Ma}, {Macleod}, {Mastropietro}, {McCully}, {Montagnac}, {Morris}, {Mueller}, {Mumford}, {Muna}, {Murphy}, {Nelson}, {Nguyen}, {Ninan}, {N{\"o}the}, {Ogaz}, {Oh}, {Parejko}, {Parley}, {Pascual}, {Patil}, {Patil}, {Plunkett}, {Prochaska}, {Rastogi}, {Reddy Janga}, {Sabater}, {Sakurikar}, {Seifert}, {Sherbert}, {Sherwood-Taylor}, {Shih}, {Sick}, {Silbiger}, {Singanamalla}, {Singer}, {Sladen}, {Sooley}, {Sornarajah}, {Streicher}, {Teuben}, {Thomas}, {Tremblay}, {Turner}, {Terr{\'o}n}, {van Kerkwijk}, {de la Vega}, {Watkins}, {Weaver}, {Whitmore}, {Woillez}, {Zabalza}, \& {Astropy Contributors}}]{astropy2018}
{Astropy Collaboration}, {Price-Whelan}, A.~M., {Sip{\H{o}}cz}, B.~M., {et~al.} 2018, \aj, 156, 123, \dodoi{10.3847/1538-3881/aabc4f}

\bibitem[{{Astropy Collaboration} {et~al.}(2022){Astropy Collaboration}, {Price-Whelan}, {Lim}, {Earl}, {Starkman}, {Bradley}, {Shupe}, {Patil}, {Corrales}, {Brasseur}, {N{\"o}the}, {Donath}, {Tollerud}, {Morris}, {Ginsburg}, {Vaher}, {Weaver}, {Tocknell}, {Jamieson}, {van Kerkwijk}, {Robitaille}, {Merry}, {Bachetti}, {G{\"u}nther}, {Aldcroft}, {Alvarado-Montes}, {Archibald}, {B{\'o}di}, {Bapat}, {Barentsen}, {Baz{\'a}n}, {Biswas}, {Boquien}, {Burke}, {Cara}, {Cara}, {Conroy}, {Conseil}, {Craig}, {Cross}, {Cruz}, {D'Eugenio}, {Dencheva}, {Devillepoix}, {Dietrich}, {Eigenbrot}, {Erben}, {Ferreira}, {Foreman-Mackey}, {Fox}, {Freij}, {Garg}, {Geda}, {Glattly}, {Gondhalekar}, {Gordon}, {Grant}, {Greenfield}, {Groener}, {Guest}, {Gurovich}, {Handberg}, {Hart}, {Hatfield-Dodds}, {Homeier}, {Hosseinzadeh}, {Jenness}, {Jones}, {Joseph}, {Kalmbach}, {Karamehmetoglu}, {Ka{\l}uszy{\'n}ski}, {Kelley}, {Kern}, {Kerzendorf}, {Koch}, {Kulumani}, {Lee}, {Ly}, {Ma}, {MacBride}, {Maljaars}, {Muna}, {Murphy}, {Norman},
  {O'Steen}, {Oman}, {Pacifici}, {Pascual}, {Pascual-Granado}, {Patil}, {Perren}, {Pickering}, {Rastogi}, {Roulston}, {Ryan}, {Rykoff}, {Sabater}, {Sakurikar}, {Salgado}, {Sanghi}, {Saunders}, {Savchenko}, {Schwardt}, {Seifert-Eckert}, {Shih}, {Jain}, {Shukla}, {Sick}, {Simpson}, {Singanamalla}, {Singer}, {Singhal}, {Sinha}, {Sip{\H{o}}cz}, {Spitler}, {Stansby}, {Streicher}, {{\v{S}}umak}, {Swinbank}, {Taranu}, {Tewary}, {Tremblay}, {de Val-Borro}, {Van Kooten}, {Vasovi{\'c}}, {Verma}, {de Miranda Cardoso}, {Williams}, {Wilson}, {Winkel}, {Wood-Vasey}, {Xue}, {Yoachim}, {Zhang}, {Zonca}, \& {Astropy Project Contributors}}]{astropy2022}
{Astropy Collaboration}, {Price-Whelan}, A.~M., {Lim}, P.~L., {et~al.} 2022, \apj, 935, 167, \dodoi{10.3847/1538-4357/ac7c74}

\bibitem[{{Barro} {et~al.}(2019){Barro}, {P{\'e}rez-Gonz{\'a}lez}, {Cava}, {Brammer}, {Pandya}, {Eliche Moral}, {Esquej}, {Dom{\'\i}nguez-S{\'a}nchez}, {Alcalde Pampliega}, {Guo}, {Koekemoer}, {Trump}, {Ashby}, {Cardiel}, {Castellano}, {Conselice}, {Dickinson}, {Dolch}, {Donley}, {Espino Briones}, {Faber}, {Fazio}, {Ferguson}, {Finkelstein}, {Fontana}, {Galametz}, {Gardner}, {Gawiser}, {Giavalisco}, {Grazian}, {Grogin}, {Hathi}, {Hemmati}, {Hern{\'a}n-Caballero}, {Kocevski}, {Koo}, {Kodra}, {Lee}, {Lin}, {Lucas}, {Mobasher}, {McGrath}, {Nandra}, {Nayyeri}, {Newman}, {Pforr}, {Peth}, {Rafelski}, {Rodr{\'\i}guez-Munoz}, {Salvato}, {Stefanon}, {van der Wel}, {Willner}, {Wiklind}, \& {Wuyts}}]{Barro2019}
{Barro}, G., {P{\'e}rez-Gonz{\'a}lez}, P.~G., {Cava}, A., {et~al.} 2019, \apjs, 243, 22, \dodoi{10.3847/1538-4365/ab23f2}

\bibitem[{{Bushouse} {et~al.}(2024){Bushouse}, Eisenhamer, Dencheva, Davies, Greenfield, Morrison, Hodge, Simon, Grumm, Droettboom, Slavich, Sosey, Pauly, Miller, Jedrzejewski, Hack, Davis, Crawford, Law, Gordon, Regan, Cara, MacDonald, Bradley, Shanahan, Jamieson, Teodoro, Williams, Pena-Guerrero, Graham, Molter, Brandt, Hayes, Cooper, \& Clarke}]{https://doi.org/10.5281/zenodo.6984365}
{Bushouse}, H., Eisenhamer, J., Dencheva, N., {et~al.} 2024, JWST Calibration Pipeline,  Zenodo, \dodoi{10.5281/ZENODO.6984365}

\bibitem[{{D'Eugenio} {et~al.}(2023){D'Eugenio}, {Perez-Gonzalez}, {Maiolino}, {Scholtz}, {Perna}, {Circosta}, {Uebler}, {Arribas}, {Boeker}, {Bunker}, {Carniani}, {Charlot}, {Chevallard}, {Cresci}, {Curtis-Lake}, {Jones}, {Kumari}, {Lamperti}, {Looser}, {Parlanti}, {Rix}, {Robertson}, {Rodriguez Del Pino}, {Tacchella}, {Venturi}, \& {Willott}}]{D'Eugenio2023}
{D'Eugenio}, F., {Perez-Gonzalez}, P., {Maiolino}, R., {et~al.} 2023, arXiv e-prints, arXiv:2308.06317, \dodoi{10.48550/arXiv.2308.06317}

\bibitem[{{Ding} {et~al.}(2020){Ding}, {Silverman}, {Treu}, {Schulze}, {Schramm}, {Birrer}, {Park}, {Jahnke}, {Bennert}, {Kartaltepe}, {Koekemoer}, {Malkan}, \& {Sanders}}]{Ding2020}
{Ding}, X., {Silverman}, J., {Treu}, T., {et~al.} 2020, \apj, 888, 37, \dodoi{10.3847/1538-4357/ab5b90}

\bibitem[{{Ding} {et~al.}(2023){Ding}, {Onoue}, {Silverman}, {Matsuoka}, {Izumi}, {Strauss}, {Jahnke}, {Phillips}, {Li}, {Volonteri}, {Haiman}, {Andika}, {Aoki}, {Baba}, {Bieri}, {Bosman}, {Bottrell}, {Eilers}, {Fujimoto}, {Habouzit}, {Imanishi}, {Inayoshi}, {Iwasawa}, {Kashikawa}, {Kawaguchi}, {Kohno}, {Lee}, {Lupi}, {Lyu}, {Nagao}, {Overzier}, {Schindler}, {Schramm}, {Shimasaku}, {Toba}, {Trakhtenbrot}, {Trebitsch}, {Treu}, {Umehata}, {Venemans}, {Vestergaard}, {Walter}, {Wang}, \& {Yang}}]{Ding2023}
{Ding}, X., {Onoue}, M., {Silverman}, J.~D., {et~al.} 2023, \nat, 621, 51, \dodoi{10.1038/s41586-023-06345-5}

\bibitem[{{Duras} {et~al.}(2020){Duras}, {Bongiorno}, {Ricci}, {Piconcelli}, {Shankar}, {Lusso}, {Bianchi}, {Fiore}, {Maiolino}, {Marconi}, {Onori}, {Sani}, {Schneider}, {Vignali}, \& {La Franca}}]{Duras2020}
{Duras}, F., {Bongiorno}, A., {Ricci}, F., {et~al.} 2020, \aap, 636, A73, \dodoi{10.1051/0004-6361/201936817}

\bibitem[{{Eisenstein} {et~al.}(2023){Eisenstein}, {Willott}, {Alberts}, {Arribas}, {Bonaventura}, {Bunker}, {Cameron}, {Carniani}, {Charlot}, {Curtis-Lake}, {D'Eugenio}, {Endsley}, {Ferruit}, {Giardino}, {Hainline}, {Hausen}, {Jakobsen}, {Johnson}, {Maiolino}, {Rieke}, {Rieke}, {Rix}, {Robertson}, {Stark}, {Tacchella}, {Williams}, {Willmer}, {Baker}, {Baum}, {Bhatawdekar}, {Boyett}, {Chen}, {Chevallard}, {Circosta}, {Curti}, {Danhaive}, {DeCoursey}, {de Graaff}, {Dressler}, {Egami}, {Helton}, {Hviding}, {Ji}, {Jones}, {Kumari}, {L{\"u}tzgendorf}, {Laseter}, {Looser}, {Lyu}, {Maseda}, {Nelson}, {Parlanti}, {Perna}, {Pusk{\'a}s}, {Rawle}, {Rodr{\'\i}guez Del Pino}, {Sandles}, {Saxena}, {Scholtz}, {Sharpe}, {Shivaei}, {Silcock}, {Simmonds}, {Skarbinski}, {Smit}, {Stone}, {Suess}, {Sun}, {Tang}, {Topping}, {{\"U}bler}, {Villanueva}, {Wallace}, {Whitler}, {Witstok}, \& {Woodrum}}]{Eisenstein2023}
{Eisenstein}, D.~J., {Willott}, C., {Alberts}, S., {et~al.} 2023, arXiv e-prints, arXiv:2306.02465, \dodoi{10.48550/arXiv.2306.02465}

\bibitem[{{Fabian}(2012)}]{Fabian2012}
{Fabian}, A.~C. 2012, \araa, 50, 455, \dodoi{10.1146/annurev-astro-081811-125521}

\bibitem[{{Ferrarese} \& {Merritt}(2000)}]{Ferrarese2000}
{Ferrarese}, L., \& {Merritt}, D. 2000, \apjl, 539, L9, \dodoi{10.1086/312838}

\bibitem[{{Gaia Collaboration} {et~al.}(2023){Gaia Collaboration}, {Vallenari}, {Brown}, {Prusti}, {de Bruijne}, {Arenou}, {Babusiaux}, {Biermann}, {Creevey}, {Ducourant}, {Evans}, {Eyer}, {Guerra}, {Hutton}, {Jordi}, {Klioner}, {Lammers}, {Lindegren}, {Luri}, {Mignard}, {Panem}, {Pourbaix}, {Randich}, {Sartoretti}, {Soubiran}, {Tanga}, {Walton}, {Bailer-Jones}, {Bastian}, {Drimmel}, {Jansen}, {Katz}, {Lattanzi}, {van Leeuwen}, {Bakker}, {Cacciari}, {Casta{\~n}eda}, {De Angeli}, {Fabricius}, {Fouesneau}, {Fr{\'e}mat}, {Galluccio}, {Guerrier}, {Heiter}, {Masana}, {Messineo}, {Mowlavi}, {Nicolas}, {Nienartowicz}, {Pailler}, {Panuzzo}, {Riclet}, {Roux}, {Seabroke}, {Sordo}, {Th{\'e}venin}, {Gracia-Abril}, {Portell}, {Teyssier}, {Altmann}, {Andrae}, {Audard}, {Bellas-Velidis}, {Benson}, {Berthier}, {Blomme}, {Burgess}, {Busonero}, {Busso}, {C{\'a}novas}, {Carry}, {Cellino}, {Cheek}, {Clementini}, {Damerdji}, {Davidson}, {de Teodoro}, {Nu{\~n}ez Campos}, {Delchambre}, {Dell'Oro}, {Esquej},
  {Fern{\'a}ndez-Hern{\'a}ndez}, {Fraile}, {Garabato}, {Garc{\'\i}a-Lario}, {Gosset}, {Haigron}, {Halbwachs}, {Hambly}, {Harrison}, {Hern{\'a}ndez}, {Hestroffer}, {Hodgkin}, {Holl}, {Jan{\ss}en}, {Jevardat de Fombelle}, {Jordan}, {Krone-Martins}, {Lanzafame}, {L{\"o}ffler}, {Marchal}, {Marrese}, {Moitinho}, {Muinonen}, {Osborne}, {Pancino}, {Pauwels}, {Recio-Blanco}, {Reyl{\'e}}, {Riello}, {Rimoldini}, {Roegiers}, {Rybizki}, {Sarro}, {Siopis}, {Smith}, {Sozzetti}, {Utrilla}, {van Leeuwen}, {Abbas}, {{\'A}brah{\'a}m}, {Abreu Aramburu}, {Aerts}, {Aguado}, {Ajaj}, {Aldea-Montero}, {Altavilla}, {{\'A}lvarez}, {Alves}, {Anders}, {Anderson}, {Anglada Varela}, {Antoja}, {Baines}, {Baker}, {Balaguer-N{\'u}{\~n}ez}, {Balbinot}, {Balog}, {Barache}, {Barbato}, {Barros}, {Barstow}, {Bartolom{\'e}}, {Bassilana}, {Bauchet}, {Becciani}, {Bellazzini}, {Berihuete}, {Bernet}, {Bertone}, {Bianchi}, {Binnenfeld}, {Blanco-Cuaresma}, {Blazere}, {Boch}, {Bombrun}, {Bossini}, {Bouquillon}, {Bragaglia}, {Bramante}, {Breedt},
  {Bressan}, {Brouillet}, {Brugaletta}, {Bucciarelli}, {Burlacu}, {Butkevich}, {Buzzi}, {Caffau}, {Cancelliere}, {Cantat-Gaudin}, {Carballo}, {Carlucci}, {Carnerero}, {Carrasco}, {Casamiquela}, {Castellani}, {Castro-Ginard}, {Chaoul}, {Charlot}, {Chemin}, {Chiaramida}, {Chiavassa}, {Chornay}, {Comoretto}, {Contursi}, {Cooper}, {Cornez}, {Cowell}, {Crifo}, {Cropper}, {Crosta}, {Crowley}, {Dafonte}, {Dapergolas}, {David}, {David}, {de Laverny}, {De Luise}, {De March}, {De Ridder}, {de Souza}, {de Torres}, {del Peloso}, {del Pozo}, {Delbo}, {Delgado}, {Delisle}, {Demouchy}, {Dharmawardena}, {Di Matteo}, {Diakite}, {Diener}, {Distefano}, {Dolding}, {Edvardsson}, {Enke}, {Fabre}, {Fabrizio}, {Faigler}, {Fedorets}, {Fernique}, {Fienga}, {Figueras}, {Fournier}, {Fouron}, {Fragkoudi}, {Gai}, {Garcia-Gutierrez}, {Garcia-Reinaldos}, {Garc{\'\i}a-Torres}, {Garofalo}, {Gavel}, {Gavras}, {Gerlach}, {Geyer}, {Giacobbe}, {Gilmore}, {Girona}, {Giuffrida}, {Gomel}, {Gomez}, {Gonz{\'a}lez-N{\'u}{\~n}ez},
  {Gonz{\'a}lez-Santamar{\'\i}a}, {Gonz{\'a}lez-Vidal}, {Granvik}, {Guillout}, {Guiraud}, {Guti{\'e}rrez-S{\'a}nchez}, {Guy}, {Hatzidimitriou}, {Hauser}, {Haywood}, {Helmer}, {Helmi}, {Sarmiento}, {Hidalgo}, {Hilger}, {H{\l}adczuk}, {Hobbs}, {Holland}, {Huckle}, {Jardine}, {Jasniewicz}, {Jean-Antoine Piccolo}, {Jim{\'e}nez-Arranz}, {Jorissen}, {Juaristi Campillo}, {Julbe}, {Karbevska}, {Kervella}, {Khanna}, {Kontizas}, {Kordopatis}, {Korn}, {K{\'o}sp{\'a}l}, {Kostrzewa-Rutkowska}, {Kruszy{\'n}ska}, {Kun}, {Laizeau}, {Lambert}, {Lanza}, {Lasne}, {Le Campion}, {Lebreton}, {Lebzelter}, {Leccia}, {Leclerc}, {Lecoeur-Taibi}, {Liao}, {Licata}, {Lindstr{\o}m}, {Lister}, {Livanou}, {Lobel}, {Lorca}, {Loup}, {Madrero Pardo}, {Magdaleno Romeo}, {Managau}, {Mann}, {Manteiga}, {Marchant}, {Marconi}, {Marcos}, {Marcos Santos}, {Mar{\'\i}n Pina}, {Marinoni}, {Marocco}, {Marshall}, {Martin Polo}, {Mart{\'\i}n-Fleitas}, {Marton}, {Mary}, {Masip}, {Massari}, {Mastrobuono-Battisti}, {Mazeh}, {McMillan}, {Messina}, {Michalik},
  {Millar}, {Mints}, {Molina}, {Molinaro}, {Moln{\'a}r}, {Monari}, {Mongui{\'o}}, {Montegriffo}, {Montero}, {Mor}, {Mora}, {Morbidelli}, {Morel}, {Morris}, {Muraveva}, {Murphy}, {Musella}, {Nagy}, {Noval}, {Oca{\~n}a}, {Ogden}, {Ordenovic}, {Osinde}, {Pagani}, {Pagano}, {Palaversa}, {Palicio}, {Pallas-Quintela}, {Panahi}, {Payne-Wardenaar}, {Pe{\~n}alosa Esteller}, {Penttil{\"a}}, {Pichon}, {Piersimoni}, {Pineau}, {Plachy}, {Plum}, {Poggio}, {Pr{\v{s}}a}, {Pulone}, {Racero}, {Ragaini}, {Rainer}, {Raiteri}, {Rambaux}, {Ramos}, {Ramos-Lerate}, {Re Fiorentin}, {Regibo}, {Richards}, {Rios Diaz}, {Ripepi}, {Riva}, {Rix}, {Rixon}, {Robichon}, {Robin}, {Robin}, {Roelens}, {Rogues}, {Rohrbasser}, {Romero-G{\'o}mez}, {Rowell}, {Royer}, {Ruz Mieres}, {Rybicki}, {Sadowski}, {S{\'a}ez N{\'u}{\~n}ez}, {Sagrist{\`a} Sell{\'e}s}, {Sahlmann}, {Salguero}, {Samaras}, {Sanchez Gimenez}, {Sanna}, {Santove{\~n}a}, {Sarasso}, {Schultheis}, {Sciacca}, {Segol}, {Segovia}, {S{\'e}gransan}, {Semeux}, {Shahaf}, {Siddiqui}, {Siebert},
  {Siltala}, {Silvelo}, {Slezak}, {Slezak}, {Smart}, {Snaith}, {Solano}, {Solitro}, {Souami}, {Souchay}, {Spagna}, {Spina}, {Spoto}, {Steele}, {Steidelm{\"u}ller}, {Stephenson}, {S{\"u}veges}, {Surdej}, {Szabados}, {Szegedi-Elek}, {Taris}, {Taylor}, {Teixeira}, {Tolomei}, {Tonello}, {Torra}, {Torra}, {Torralba Elipe}, {Trabucchi}, {Tsounis}, {Turon}, {Ulla}, {Unger}, {Vaillant}, {van Dillen}, {van Reeven}, {Vanel}, {Vecchiato}, {Viala}, {Vicente}, {Voutsinas}, {Weiler}, {Wevers}, {Wyrzykowski}, {Yoldas}, {Yvard}, {Zhao}, {Zorec}, {Zucker}, \& {Zwitter}}]{Gaia2023}
{Gaia Collaboration}, {Vallenari}, A., {Brown}, A.~G.~A., {et~al.} 2023, \aap, 674, A1, \dodoi{10.1051/0004-6361/202243940}

\bibitem[{{Gebhardt} {et~al.}(2000){Gebhardt}, {Bender}, {Bower}, {Dressler}, {Faber}, {Filippenko}, {Green}, {Grillmair}, {Ho}, {Kormendy}, {Lauer}, {Magorrian}, {Pinkney}, {Richstone}, \& {Tremaine}}]{Gebhardt2000}
{Gebhardt}, K., {Bender}, R., {Bower}, G., {et~al.} 2000, \apjl, 539, L13, \dodoi{10.1086/312840}

\bibitem[{{Giavalisco} {et~al.}(2004){Giavalisco}, {Ferguson}, {Koekemoer}, {Dickinson}, {Alexander}, {Bauer}, {Bergeron}, {Biagetti}, {Brandt}, {Casertano}, {Cesarsky}, {Chatzichristou}, {Conselice}, {Cristiani}, {Da Costa}, {Dahlen}, {de Mello}, {Eisenhardt}, {Erben}, {Fall}, {Fassnacht}, {Fosbury}, {Fruchter}, {Gardner}, {Grogin}, {Hook}, {Hornschemeier}, {Idzi}, {Jogee}, {Kretchmer}, {Laidler}, {Lee}, {Livio}, {Lucas}, {Madau}, {Mobasher}, {Moustakas}, {Nonino}, {Padovani}, {Papovich}, {Park}, {Ravindranath}, {Renzini}, {Richardson}, {Riess}, {Rosati}, {Schirmer}, {Schreier}, {Somerville}, {Spinrad}, {Stern}, {Stiavelli}, {Strolger}, {Urry}, {Vandame}, {Williams}, \& {Wolf}}]{Giavalisco2004}
{Giavalisco}, M., {Ferguson}, H.~C., {Koekemoer}, A.~M., {et~al.} 2004, \apjl, 600, L93, \dodoi{10.1086/379232}

\bibitem[{{Greene} \& {Ho}(2005)}]{Greene2005}
{Greene}, J.~E., \& {Ho}, L.~C. 2005, \apj, 630, 122, \dodoi{10.1086/431897}

\bibitem[{{Greene} {et~al.}(2020){Greene}, {Strader}, \& {Ho}}]{Greene2020}
{Greene}, J.~E., {Strader}, J., \& {Ho}, L.~C. 2020, \araa, 58, 257, \dodoi{10.1146/annurev-astro-032620-021835}

\bibitem[{{Harikane} {et~al.}(2023){Harikane}, {Zhang}, {Nakajima}, {Ouchi}, {Isobe}, {Ono}, {Hatano}, {Xu}, \& {Umeda}}]{Harikane2023}
{Harikane}, Y., {Zhang}, Y., {Nakajima}, K., {et~al.} 2023, \apj, 959, 39, \dodoi{10.3847/1538-4357/ad029e}

\bibitem[{{H{\"a}ring} \& {Rix}(2004)}]{Haring2004}
{H{\"a}ring}, N., \& {Rix}, H.-W. 2004, \apjl, 604, L89, \dodoi{10.1086/383567}

\bibitem[{{Harrison}(2017)}]{Harrison2017}
{Harrison}, C.~M. 2017, Nature Astronomy, 1, 0165, \dodoi{10.1038/s41550-017-0165}

\bibitem[{{Hazard} {et~al.}(1963){Hazard}, {Mackey}, \& {Shimmins}}]{Hazard1963}
{Hazard}, C., {Mackey}, M.~B., \& {Shimmins}, A.~J. 1963, \nat, 197, 1037, \dodoi{10.1038/1971037a0}

\bibitem[{{Heckman} \& {Best}(2014)}]{Heckman2014}
{Heckman}, T.~M., \& {Best}, P.~N. 2014, \araa, 52, 589, \dodoi{10.1146/annurev-astro-081913-035722}

\bibitem[{{Hopkins} {et~al.}(2008{\natexlab{a}}){Hopkins}, {Cox}, {Kere{\v{s}}}, \& {Hernquist}}]{Hopkins2008b}
{Hopkins}, P.~F., {Cox}, T.~J., {Kere{\v{s}}}, D., \& {Hernquist}, L. 2008{\natexlab{a}}, \apjs, 175, 390, \dodoi{10.1086/524363}

\bibitem[{{Hopkins} {et~al.}(2008{\natexlab{b}}){Hopkins}, {Hernquist}, {Cox}, \& {Kere{\v{s}}}}]{Hopkins2008}
{Hopkins}, P.~F., {Hernquist}, L., {Cox}, T.~J., \& {Kere{\v{s}}}, D. 2008{\natexlab{b}}, \apjs, 175, 356, \dodoi{10.1086/524362}

\bibitem[{{Horne}(1986)}]{Horne1986}
{Horne}, K. 1986, \pasp, 98, 609, \dodoi{10.1086/131801}

\bibitem[{{Jahnke} \& {Macci{\`o}}(2011)}]{Jahnke2011}
{Jahnke}, K., \& {Macci{\`o}}, A.~V. 2011, \apj, 734, 92, \dodoi{10.1088/0004-637X/734/2/92}

\bibitem[{{Ji} {et~al.}(2023){Ji}, {Williams}, {Tacchella}, {Suess}, {Baker}, {Alberts}, {Bunker}, {Johnson}, {Robertson}, {Sun}, {Eisenstein}, {Rieke}, {Maseda}, {Hainline}, {Hausen}, {Rieke}, {Willmer}, {Egami}, {Shivaei}, {Carniani}, {Charlot}, {Chevallard}, {Curtis-Lake}, {Looser}, {Maiolino}, {Willott}, {Chen}, {Helton}, {Lyu}, {Nelson}, {Bhatawdekar}, {Boyett}, \& {Sandles}}]{Ji2023}
{Ji}, Z., {Williams}, C.~C., {Tacchella}, S., {et~al.} 2023, arXiv e-prints, arXiv:2305.18518, \dodoi{10.48550/arXiv.2305.18518}

\bibitem[{{Johnson} {et~al.}(2021){Johnson}, {Leja}, {Conroy}, \& {Speagle}}]{Johnson2021}
{Johnson}, B.~D., {Leja}, J., {Conroy}, C., \& {Speagle}, J.~S. 2021, \apjs, 254, 22, \dodoi{10.3847/1538-4365/abef67}

\bibitem[{{Kelly} \& {Shen}(2013)}]{KS2013}
{Kelly}, B.~C., \& {Shen}, Y. 2013, \apj, 764, 45, \dodoi{10.1088/0004-637X/764/1/45}

\bibitem[{Kennedy \& Eberhart(1995)}]{Kennedy1995}
Kennedy, J., \& Eberhart, R. 1995, in Proceedings of ICNN'95 - International Conference on Neural Networks, Vol.~4, 1942--1948 vol.4, \dodoi{10.1109/ICNN.1995.488968}

\bibitem[{{Koekemoer} {et~al.}(2011){Koekemoer}, {Faber}, {Ferguson}, {Grogin}, {Kocevski}, {Koo}, {Lai}, {Lotz}, {Lucas}, {McGrath}, {Ogaz}, {Rajan}, {Riess}, {Rodney}, {Strolger}, {Casertano}, {Castellano}, {Dahlen}, {Dickinson}, {Dolch}, {Fontana}, {Giavalisco}, {Grazian}, {Guo}, {Hathi}, {Huang}, {van der Wel}, {Yan}, {Acquaviva}, {Alexander}, {Almaini}, {Ashby}, {Barden}, {Bell}, {Bournaud}, {Brown}, {Caputi}, {Cassata}, {Challis}, {Chary}, {Cheung}, {Cirasuolo}, {Conselice}, {Roshan Cooray}, {Croton}, {Daddi}, {Dav{\'e}}, {de Mello}, {de Ravel}, {Dekel}, {Donley}, {Dunlop}, {Dutton}, {Elbaz}, {Fazio}, {Filippenko}, {Finkelstein}, {Frazer}, {Gardner}, {Garnavich}, {Gawiser}, {Gruetzbauch}, {Hartley}, {H{\"a}ussler}, {Herrington}, {Hopkins}, {Huang}, {Jha}, {Johnson}, {Kartaltepe}, {Khostovan}, {Kirshner}, {Lani}, {Lee}, {Li}, {Madau}, {McCarthy}, {McIntosh}, {McLure}, {McPartland}, {Mobasher}, {Moreira}, {Mortlock}, {Moustakas}, {Mozena}, {Nandra}, {Newman}, {Nielsen}, {Niemi}, {Noeske}, {Papovich},
  {Pentericci}, {Pope}, {Primack}, {Ravindranath}, {Reddy}, {Renzini}, {Rix}, {Robaina}, {Rosario}, {Rosati}, {Salimbeni}, {Scarlata}, {Siana}, {Simard}, {Smidt}, {Snyder}, {Somerville}, {Spinrad}, {Straughn}, {Telford}, {Teplitz}, {Trump}, {Vargas}, {Villforth}, {Wagner}, {Wandro}, {Wechsler}, {Weiner}, {Wiklind}, {Wild}, {Wilson}, {Wuyts}, \& {Yun}}]{Koekemoer2011}
{Koekemoer}, A.~M., {Faber}, S.~M., {Ferguson}, H.~C., {et~al.} 2011, \apjs, 197, 36, \dodoi{10.1088/0067-0049/197/2/36}

\bibitem[{{Kormendy} \& {Ho}(2013)}]{Kormendy2013}
{Kormendy}, J., \& {Ho}, L.~C. 2013, \araa, 51, 511, \dodoi{10.1146/annurev-astro-082708-101811}

\bibitem[{{Kriek} \& {Conroy}(2013)}]{Kriek2013}
{Kriek}, M., \& {Conroy}, C. 2013, \apjl, 775, L16, \dodoi{10.1088/2041-8205/775/1/L16}

\bibitem[{{Krywult} {et~al.}(2017){Krywult}, {Tasca}, {Pollo}, {Vergani}, {Bolzonella}, {Davidzon}, {Iovino}, {Gargiulo}, {Haines}, {Scodeggio}, {Guzzo}, {Zamorani}, {Garilli}, {Granett}, {de la Torre}, {Abbas}, {Adami}, {Bottini}, {Cappi}, {Cucciati}, {Franzetti}, {Fritz}, {Le Brun}, {Le F{\`e}vre}, {Maccagni}, {Ma{\l}ek}, {Marulli}, {Polletta}, {Tojeiro}, {Zanichelli}, {Arnouts}, {Bel}, {Branchini}, {Coupon}, {De Lucia}, {Ilbert}, {McCracken}, {Moscardini}, \& {Takeuchi}}]{Krywult2017}
{Krywult}, J., {Tasca}, L.~A.~M., {Pollo}, A., {et~al.} 2017, \aap, 598, A120, \dodoi{10.1051/0004-6361/201628953}

\bibitem[{{Kuhn} {et~al.}(2024){Kuhn}, {Shangguan}, {Davies}, {Man}, {Cao}, {Dexter}, {Eisenhauer}, {F{\"o}rster Schreiber}, {Feuchtgruber}, {Genzel}, {Gillessen}, {H{\"o}nig}, {Lutz}, {Netzer}, {Ott}, {Rabien}, {Santos}, {Shimizu}, {Sturm}, \& {Tacconi}}]{Kuhn2024}
{Kuhn}, L., {Shangguan}, J., {Davies}, R., {et~al.} 2024, \aap, 684, A52, \dodoi{10.1051/0004-6361/202348138}

\bibitem[{{Landt} {et~al.}(2008){Landt}, {Bentz}, {Ward}, {Elvis}, {Peterson}, {Korista}, \& {Karovska}}]{Landt2008}
{Landt}, H., {Bentz}, M.~C., {Ward}, M.~J., {et~al.} 2008, \apjs, 174, 282, \dodoi{10.1086/522373}

\bibitem[{{Lauer} {et~al.}(2007){Lauer}, {Tremaine}, {Richstone}, \& {Faber}}]{Lauer2007}
{Lauer}, T.~R., {Tremaine}, S., {Richstone}, D., \& {Faber}, S.~M. 2007, \apj, 670, 249, \dodoi{10.1086/522083}

\bibitem[{{Le} {et~al.}(2020){Le}, {Woo}, \& {Xue}}]{Le2020}
{Le}, H. A.~N., {Woo}, J.-H., \& {Xue}, Y. 2020, \apj, 901, 35, \dodoi{10.3847/1538-4357/abada0}

\bibitem[{{Li} {et~al.}(2021){Li}, {Silverman}, {Ding}, {Strauss}, {Goulding}, {Schramm}, {Yesuf}, {Sun}, {Xue}, {Birrer}, {Shi}, {Toba}, {Nagao}, \& {Imanishi}}]{Li2021}
{Li}, J., {Silverman}, J.~D., {Ding}, X., {et~al.} 2021, \apj, 922, 142, \dodoi{10.3847/1538-4357/ac2301}

\bibitem[{{Li} {et~al.}(2024){Li}, {Silverman}, {Shen}, {Volonteri}, {Jahnke}, {Zhuang}, {Scoggins}, {Ding}, {Harikane}, {Onoue}, \& {Tanaka}}]{Li2024}
{Li}, J., {Silverman}, J.~D., {Shen}, Y., {et~al.} 2024, arXiv e-prints, arXiv:2403.00074, \dodoi{10.48550/arXiv.2403.00074}

\bibitem[{{Li} {et~al.}(2023){Li}, {Shen}, {Ho}, {Brandt}, {Grier}, {Hall}, {Homayouni}, {Koekemoer}, {Schneider}, \& {Trump}}]{JLi2023}
{Li}, J. I.~H., {Shen}, Y., {Ho}, L.~C., {et~al.} 2023, \apj, 954, 173, \dodoi{10.3847/1538-4357/acddda}

\bibitem[{{Luo} {et~al.}(2017){Luo}, {Brandt}, {Xue}, {Lehmer}, {Alexander}, {Bauer}, {Vito}, {Yang}, {Basu-Zych}, {Comastri}, {Gilli}, {Gu}, {Hornschemeier}, {Koekemoer}, {Liu}, {Mainieri}, {Paolillo}, {Ranalli}, {Rosati}, {Schneider}, {Shemmer}, {Smail}, {Sun}, {Tozzi}, {Vignali}, \& {Wang}}]{Luo2017}
{Luo}, B., {Brandt}, W.~N., {Xue}, Y.~Q., {et~al.} 2017, \apjs, 228, 2, \dodoi{10.3847/1538-4365/228/1/2}

\bibitem[{{Lyu} {et~al.}(2022){Lyu}, {Alberts}, {Rieke}, \& {Rujopakarn}}]{Lyu2022}
{Lyu}, J., {Alberts}, S., {Rieke}, G.~H., \& {Rujopakarn}, W. 2022, \apj, 941, 191, \dodoi{10.3847/1538-4357/ac9e5d}

\bibitem[{{Lyu} {et~al.}(2024){Lyu}, {Alberts}, {Rieke}, {Shivaei}, {P{\'e}rez-Gonz{\'a}lez}, {Sun}, {Hainline}, {Baum}, {Bonaventura}, {Bunker}, {Egami}, {Eisenstein}, {Florian}, {Ji}, {Johnson}, {Morrison}, {Rieke}, {Robertson}, {Rujopakarn}, {Tacchella}, {Scholtz}, \& {Willmer}}]{Lyu2024}
{Lyu}, J., {Alberts}, S., {Rieke}, G.~H., {et~al.} 2024, \apj, 966, 229, \dodoi{10.3847/1538-4357/ad3643}

\bibitem[{{Magorrian} {et~al.}(1998){Magorrian}, {Tremaine}, {Richstone}, {Bender}, {Bower}, {Dressler}, {Faber}, {Gebhardt}, {Green}, {Grillmair}, {Kormendy}, \& {Lauer}}]{Magorrian1998}
{Magorrian}, J., {Tremaine}, S., {Richstone}, D., {et~al.} 1998, \aj, 115, 2285, \dodoi{10.1086/300353}

\bibitem[{{Maiolino} {et~al.}(2023){Maiolino}, {Scholtz}, {Curtis-Lake}, {Carniani}, {Baker}, {de Graaff}, {Tacchella}, {{\"U}bler}, {D'Eugenio}, {Witstok}, {Curti}, {Arribas}, {Bunker}, {Charlot}, {Chevallard}, {Eisenstein}, {Egami}, {Ji}, {Jones}, {Lyu}, {Rawle}, {Robertson}, {Rujopakarn}, {Perna}, {Sun}, {Venturi}, {Williams}, \& {Willott}}]{Maiolino2023}
{Maiolino}, R., {Scholtz}, J., {Curtis-Lake}, E., {et~al.} 2023, arXiv e-prints, arXiv:2308.01230, \dodoi{10.48550/arXiv.2308.01230}

\bibitem[{{Matthee} {et~al.}(2023){Matthee}, {Naidu}, {Brammer}, {Chisholm}, {Eilers}, {Goulding}, {Greene}, {Kashino}, {Labbe}, {Lilly}, {Mackenzie}, {Oesch}, {Weibel}, {Wuyts}, {Xiao}, {Bordoloi}, {Bouwens}, {van Dokkum}, {Illingworth}, {Kramarenko}, {Maseda}, {Mason}, {Meyer}, {Nelson}, {Reddy}, {Shivaei}, {Simcoe}, \& {Yue}}]{Matthee2023}
{Matthee}, J., {Naidu}, R.~P., {Brammer}, G., {et~al.} 2023, arXiv e-prints, arXiv:2306.05448, \dodoi{10.48550/arXiv.2306.05448}

\bibitem[{{Mechtley} {et~al.}(2016){Mechtley}, {Jahnke}, {Windhorst}, {Andrae}, {Cisternas}, {Cohen}, {Hewlett}, {Koekemoer}, {Schramm}, {Schulze}, {Silverman}, {Villforth}, {van der Wel}, \& {Wisotzki}}]{Mechtley2016}
{Mechtley}, M., {Jahnke}, K., {Windhorst}, R.~A., {et~al.} 2016, \apj, 830, 156, \dodoi{10.3847/0004-637X/830/2/156}

\bibitem[{{Merloni} {et~al.}(2010){Merloni}, {Bongiorno}, {Bolzonella}, {Brusa}, {Civano}, {Comastri}, {Elvis}, {Fiore}, {Gilli}, {Hao}, {Jahnke}, {Koekemoer}, {Lusso}, {Mainieri}, {Mignoli}, {Miyaji}, {Renzini}, {Salvato}, {Silverman}, {Trump}, {Vignali}, {Zamorani}, {Capak}, {Lilly}, {Sanders}, {Taniguchi}, {Bardelli}, {Carollo}, {Caputi}, {Contini}, {Coppa}, {Cucciati}, {de la Torre}, {de Ravel}, {Franzetti}, {Garilli}, {Hasinger}, {Impey}, {Iovino}, {Iwasawa}, {Kampczyk}, {Kneib}, {Knobel}, {Kova{\v{c}}}, {Lamareille}, {Le Borgne}, {Le Brun}, {Le F{\`e}vre}, {Maier}, {Pello}, {Peng}, {Perez Montero}, {Ricciardelli}, {Scodeggio}, {Tanaka}, {Tasca}, {Tresse}, {Vergani}, \& {Zucca}}]{Merloni2010}
{Merloni}, A., {Bongiorno}, A., {Bolzonella}, M., {et~al.} 2010, \apj, 708, 137, \dodoi{10.1088/0004-637X/708/1/137}

\bibitem[{{Mezcua} {et~al.}(2024){Mezcua}, {Pacucci}, {Suh}, {Siudek}, \& {Natarajan}}]{Mezcua2024}
{Mezcua}, M., {Pacucci}, F., {Suh}, H., {Siudek}, M., \& {Natarajan}, P. 2024, \apjl, 966, L30, \dodoi{10.3847/2041-8213/ad3c2a}

\bibitem[{{Mezcua} {et~al.}(2023){Mezcua}, {Siudek}, {Suh}, {Valiante}, {Spinoso}, \& {Bonoli}}]{Mezcua2023}
{Mezcua}, M., {Siudek}, M., {Suh}, H., {et~al.} 2023, \apjl, 943, L5, \dodoi{10.3847/2041-8213/acae25}

\bibitem[{{Mountrichas}(2023)}]{Mountrichas2023}
{Mountrichas}, G. 2023, \aap, 672, A98, \dodoi{10.1051/0004-6361/202345924}

\bibitem[{{Newville} {et~al.}(2014){Newville}, {Stensitzki}, {Allen}, \& {Ingargiola}}]{Newville2014}
{Newville}, M., {Stensitzki}, T., {Allen}, D.~B., \& {Ingargiola}, A. 2014, {LMFIT: Non-Linear Least-Square Minimization and Curve-Fitting for Python}, 0.8.0,  Zenodo, \dodoi{10.5281/zenodo.11813}

\bibitem[{{Oesch} \& {Magee}(2023)}]{https://doi.org/10.17909/gdyc-7g80}
{Oesch}, P., \& {Magee}, D. 2023, The JWST FRESCO Survey,  STScI/MAST, \dodoi{10.17909/GDYC-7G80}

\bibitem[{{Oesch} {et~al.}(2023){Oesch}, {Brammer}, {Naidu}, {Bouwens}, {Chisholm}, {Illingworth}, {Matthee}, {Nelson}, {Qin}, {Reddy}, {Shapley}, {Shivaei}, {van Dokkum}, {Weibel}, {Whitaker}, {Wuyts}, {Covelo-Paz}, {Endsley}, {Fudamoto}, {Giovinazzo}, {Herard-Demanche}, {Kerutt}, {Kramarenko}, {Labbe}, {Leonova}, {Lin}, {Magee}, {Marchesini}, {Maseda}, {Mason}, {Matharu}, {Meyer}, {Neufeld}, {Prieto Lyon}, {Schaerer}, {Sharma}, {Shuntov}, {Smit}, {Stefanon}, {Wyithe}, \& {Xiao}}]{Oesch2023}
{Oesch}, P.~A., {Brammer}, G., {Naidu}, R.~P., {et~al.} 2023, \mnras, 525, 2864, \dodoi{10.1093/mnras/stad2411}

\bibitem[{{Pacucci} {et~al.}(2023){Pacucci}, {Nguyen}, {Carniani}, {Maiolino}, \& {Fan}}]{Pacucci2023}
{Pacucci}, F., {Nguyen}, B., {Carniani}, S., {Maiolino}, R., \& {Fan}, X. 2023, \apjl, 957, L3, \dodoi{10.3847/2041-8213/ad0158}

\bibitem[{{Peng}(2007)}]{Peng2007}
{Peng}, C.~Y. 2007, \apj, 671, 1098, \dodoi{10.1086/522774}

\bibitem[{{Peng} {et~al.}(2006){Peng}, {Impey}, {Rix}, {Kochanek}, {Keeton}, {Falco}, {Leh{\'a}r}, \& {McLeod}}]{Peng2006}
{Peng}, C.~Y., {Impey}, C.~D., {Rix}, H.-W., {et~al.} 2006, \apj, 649, 616, \dodoi{10.1086/506266}

\bibitem[{{Peterson}(2006)}]{Peterson2006}
{Peterson}, B.~M. 2006, in Physics of Active Galactic Nuclei at all Scales, ed. D.~{Alloin}, Vol. 693, 77, \dodoi{10.1007/3-540-34621-X_3}

\bibitem[{{Popovi{\'c}} {et~al.}(2019){Popovi{\'c}}, {Kova{\v{c}}evi{\'c}-Doj{\v{c}}inovi{\'c}}, \& {Mar{\v{c}}eta-Mandi{\'c}}}]{Popovic2019}
{Popovi{\'c}}, L.~{\v{C}}., {Kova{\v{c}}evi{\'c}-Doj{\v{c}}inovi{\'c}}, J., \& {Mar{\v{c}}eta-Mandi{\'c}}, S. 2019, \mnras, 484, 3180, \dodoi{10.1093/mnras/stz157}

\bibitem[{{Reines} \& {Volonteri}(2015)}]{Reines2015}
{Reines}, A.~E., \& {Volonteri}, M. 2015, \apj, 813, 82, \dodoi{10.1088/0004-637X/813/2/82}

\bibitem[{{Ricci} {et~al.}(2017){Ricci}, {La Franca}, {Onori}, \& {Bianchi}}]{Ricci2017}
{Ricci}, F., {La Franca}, F., {Onori}, F., \& {Bianchi}, S. 2017, \aap, 598, A51, \dodoi{10.1051/0004-6361/201629380}

\bibitem[{{Ricci} {et~al.}(2022){Ricci}, {Treister}, {Bauer}, {Mej{\'\i}a-Restrepo}, {Koss}, {den Brok}, {Balokovi{\'c}}, {B{\"a}r}, {Bessiere}, {Caglar}, {Harrison}, {Ichikawa}, {Kakkad}, {Lamperti}, {Mushotzky}, {Oh}, {Powell}, {Privon}, {Ricci}, {Riffel}, {Rojas}, {Sani}, {Smith}, {Stern}, {Trakhtenbrot}, {Urry}, \& {Veilleux}}]{Ricci2022}
{Ricci}, F., {Treister}, E., {Bauer}, F.~E., {et~al.} 2022, \apjs, 261, 8, \dodoi{10.3847/1538-4365/ac5b67}

\bibitem[{{Rieke} {et~al.}(2024){Rieke}, {Alberts}, {Shivaei}, {Lyu}, {Willmer}, {Perez-Gonzalez}, \& {Williams}}]{Rieke2024}
{Rieke}, G., {Alberts}, S., {Shivaei}, I., {et~al.} 2024, arXiv e-prints, arXiv:2406.03518, \dodoi{10.48550/arXiv.2406.03518}

\bibitem[{{Rieke} {et~al.}(2023{\natexlab{a}}){Rieke}, {Robertson}, {Tacchella}, {Willmer}, {Johnson}, {Carniani}, {Bunker}, \& {Willott}}]{https://doi.org/10.17909/8tdj-8n28}
{Rieke}, M., {Robertson}, B., {Tacchella}, S., {et~al.} 2023{\natexlab{a}}, Data from the JWST Advanced Deep Extragalactic Survey (JADES),  STScI/MAST, \dodoi{10.17909/8TDJ-8N28}

\bibitem[{{Rieke} {et~al.}(2023{\natexlab{b}}){Rieke}, {Kelly}, {Misselt}, {Stansberry}, {Boyer}, {Beatty}, {Egami}, {Florian}, {Greene}, {Hainline}, {Leisenring}, {Roellig}, {Schlawin}, {Sun}, {Tinnin}, {Williams}, {Willmer}, {Wilson}, {Clark}, {Rohrbach}, {Brooks}, {Canipe}, {Correnti}, {DiFelice}, {Gennaro}, {Girard}, {Hartig}, {Hilbert}, {Koekemoer}, {Nikolov}, {Pirzkal}, {Rest}, {Robberto}, {Sunnquist}, {Telfer}, {Wu}, {Ferry}, {Lewis}, {Baum}, {Beichman}, {Doyon}, {Dressler}, {Eisenstein}, {Ferrarese}, {Hodapp}, {Horner}, {Jaffe}, {Johnstone}, {Krist}, {Martin}, {McCarthy}, {Meyer}, {Rieke}, {Trauger}, \& {Young}}]{Rieke2023}
{Rieke}, M.~J., {Kelly}, D.~M., {Misselt}, K., {et~al.} 2023{\natexlab{b}}, \pasp, 135, 028001, \dodoi{10.1088/1538-3873/acac53}

\bibitem[{{Schmidt}(1963)}]{Schmidt1963}
{Schmidt}, M. 1963, \nat, 197, 1040, \dodoi{10.1038/1971040a0}

\bibitem[{{Schramm} \& {Silverman}(2013)}]{Schramm2013}
{Schramm}, M., \& {Silverman}, J.~D. 2013, \apj, 767, 13, \dodoi{10.1088/0004-637X/767/1/13}

\bibitem[{{Schulze} \& {Wisotzki}(2011)}]{Schulze2011}
{Schulze}, A., \& {Wisotzki}, L. 2011, \aap, 535, A87, \dodoi{10.1051/0004-6361/201117564}

\bibitem[{{Schulze} \& {Wisotzki}(2014)}]{Schulze2014}
---. 2014, \mnras, 438, 3422, \dodoi{10.1093/mnras/stt2457}

\bibitem[{{Schulze} {et~al.}(2015){Schulze}, {Bongiorno}, {Gavignaud}, {Schramm}, {Silverman}, {Merloni}, {Zamorani}, {Hirschmann}, {Mainieri}, {Wisotzki}, {Shankar}, {Fiore}, {Koekemoer}, \& {Temporin}}]{Schulze2015}
{Schulze}, A., {Bongiorno}, A., {Gavignaud}, I., {et~al.} 2015, \mnras, 447, 2085, \dodoi{10.1093/mnras/stu2549}

\bibitem[{{Setoguchi} {et~al.}(2021){Setoguchi}, {Ueda}, {Toba}, \& {Akiyama}}]{Setoguchi2021}
{Setoguchi}, K., {Ueda}, Y., {Toba}, Y., \& {Akiyama}, M. 2021, \apj, 909, 188, \dodoi{10.3847/1538-4357/abdf55}

\bibitem[{{Shen} \& {Liu}(2012)}]{Shen2012}
{Shen}, Y., \& {Liu}, X. 2012, \apj, 753, 125, \dodoi{10.1088/0004-637X/753/2/125}

\bibitem[{{Shen} {et~al.}(2016){Shen}, {Brandt}, {Richards}, {Denney}, {Greene}, {Grier}, {Ho}, {Peterson}, {Petitjean}, {Schneider}, {Tao}, \& {Trump}}]{Shen2016}
{Shen}, Y., {Brandt}, W.~N., {Richards}, G.~T., {et~al.} 2016, \apj, 831, 7, \dodoi{10.3847/0004-637X/831/1/7}

\bibitem[{{Springel} {et~al.}(2005){Springel}, {White}, {Jenkins}, {Frenk}, {Yoshida}, {Gao}, {Navarro}, {Thacker}, {Croton}, {Helly}, {Peacock}, {Cole}, {Thomas}, {Couchman}, {Evrard}, {Colberg}, \& {Pearce}}]{Springel2005}
{Springel}, V., {White}, S. D.~M., {Jenkins}, A., {et~al.} 2005, \nat, 435, 629, \dodoi{10.1038/nature03597}

\bibitem[{{Stone} {et~al.}(2024){Stone}, {Lyu}, {Rieke}, {Alberts}, \& {Hainline}}]{Stone2024}
{Stone}, M.~A., {Lyu}, J., {Rieke}, G.~H., {Alberts}, S., \& {Hainline}, K.~N. 2024, \apj, 964, 90, \dodoi{10.3847/1538-4357/ad2a57}

\bibitem[{{Suh} {et~al.}(2020){Suh}, {Civano}, {Trakhtenbrot}, {Shankar}, {Hasinger}, {Sanders}, \& {Allevato}}]{Suh2020}
{Suh}, H., {Civano}, F., {Trakhtenbrot}, B., {et~al.} 2020, \apj, 889, 32, \dodoi{10.3847/1538-4357/ab5f5f}

\bibitem[{{Sun}(2024)}]{https://doi.org/10.5281/zenodo.14052875}
{Sun}, F. 2024, nircam\_grism,  Zenodo, \dodoi{10.5281/ZENODO.14052875}

\bibitem[{{Sun} {et~al.}(2023){Sun}, {Egami}, {Pirzkal}, {Rieke}, {Baum}, {Boyer}, {Boyett}, {Bunker}, {Cameron}, {Curti}, {Eisenstein}, {Gennaro}, {Greene}, {Jaffe}, {Kelly}, {Koekemoer}, {Kumari}, {Maiolino}, {Maseda}, {Perna}, {Rest}, {Robertson}, {Schlawin}, {Smit}, {Stansberry}, {Sunnquist}, {Tacchella}, {Williams}, \& {Willmer}}]{Sun2023}
{Sun}, F., {Egami}, E., {Pirzkal}, N., {et~al.} 2023, \apj, 953, 53, \dodoi{10.3847/1538-4357/acd53c}

\bibitem[{{Sun} {et~al.}(2015){Sun}, {Trump}, {Brandt}, {Luo}, {Alexander}, {Jahnke}, {Rosario}, {Wang}, \& {Xue}}]{Sun2015}
{Sun}, M., {Trump}, J.~R., {Brandt}, W.~N., {et~al.} 2015, \apj, 802, 14, \dodoi{10.1088/0004-637X/802/1/14}

\bibitem[{{Tanaka} {et~al.}(2024){Tanaka}, {Silverman}, {Ding}, {Jahnke}, {Trakhtenbrot}, {Lambrides}, {Onoue}, {Taufik Andika}, {Bongiorno}, {Faisst}, {Gillman}, {Hayward}, {Hirschmann}, {Koekemoer}, {Kokorev}, {Liu}, {Magdis}, {Renzini}, {Casey}, {Drakos}, {Franco}, {Gozaliasl}, {Kartaltepe}, {Liu}, {McCracken}, {Rhodes}, {Robertson}, \& {Toft}}]{Tanaka2024}
{Tanaka}, T.~S., {Silverman}, J.~D., {Ding}, X., {et~al.} 2024, arXiv e-prints, arXiv:2401.13742, \dodoi{10.48550/arXiv.2401.13742}

\bibitem[{{Tremaine} {et~al.}(2002){Tremaine}, {Gebhardt}, {Bender}, {Bower}, {Dressler}, {Faber}, {Filippenko}, {Green}, {Grillmair}, {Ho}, {Kormendy}, {Lauer}, {Magorrian}, {Pinkney}, \& {Richstone}}]{Tremaine2002}
{Tremaine}, S., {Gebhardt}, K., {Bender}, R., {et~al.} 2002, \apj, 574, 740, \dodoi{10.1086/341002}

\bibitem[{{{\"U}bler} {et~al.}(2023){{\"U}bler}, {Maiolino}, {Curtis-Lake}, {P{\'e}rez-Gonz{\'a}lez}, {Curti}, {Perna}, {Arribas}, {Charlot}, {Marshall}, {D'Eugenio}, {Scholtz}, {Bunker}, {Carniani}, {Ferruit}, {Jakobsen}, {Rix}, {Rodr{\'\i}guez Del Pino}, {Willott}, {Boeker}, {Cresci}, {Jones}, {Kumari}, \& {Rawle}}]{Ubler2023}
{{\"U}bler}, H., {Maiolino}, R., {Curtis-Lake}, E., {et~al.} 2023, \aap, 677, A145, \dodoi{10.1051/0004-6361/202346137}

\bibitem[{{Vestergaard}(2002)}]{Vestergaard2002}
{Vestergaard}, M. 2002, \apj, 571, 733, \dodoi{10.1086/340045}

\bibitem[{{Virtanen} {et~al.}(2020){Virtanen}, {Gommers}, {Oliphant}, {Haberland}, {Reddy}, {Cournapeau}, {Burovski}, {Peterson}, {Weckesser}, {Bright}, {van der Walt}, {Brett}, {Wilson}, {Millman}, {Mayorov}, {Nelson}, {Jones}, {Kern}, {Larson}, {Carey}, {Polat}, {Feng}, {Moore}, {VanderPlas}, {Laxalde}, {Perktold}, {Cimrman}, {Henriksen}, {Quintero}, {Harris}, {Archibald}, {Ribeiro}, {Pedregosa}, {van Mulbregt}, \& {SciPy 1. 0 Contributors}}]{Virtanen2020}
{Virtanen}, P., {Gommers}, R., {Oliphant}, T.~E., {et~al.} 2020, Nature Methods, 17, 261, \dodoi{10.1038/s41592-019-0686-2}

\bibitem[{{Weaver} {et~al.}(2023){Weaver}, {Davidzon}, {Toft}, {Ilbert}, {McCracken}, {Gould}, {Jespersen}, {Steinhardt}, {Lagos}, {Capak}, {Casey}, {Chartab}, {Faisst}, {Hayward}, {Kartaltepe}, {Kauffmann}, {Koekemoer}, {Kokorev}, {Laigle}, {Liu}, {Long}, {Magdis}, {McPartland}, {Milvang-Jensen}, {Mobasher}, {Moneti}, {Peng}, {Sanders}, {Shuntov}, {Sneppen}, {Valentino}, {Zalesky}, \& {Zamorani}}]{Weaver2023}
{Weaver}, J.~R., {Davidzon}, I., {Toft}, S., {et~al.} 2023, \aap, 677, A184, \dodoi{10.1051/0004-6361/202245581}

\bibitem[{{Willott} {et~al.}(2005){Willott}, {Percival}, {McLure}, {Crampton}, {Hutchings}, {Jarvis}, {Sawicki}, \& {Simard}}]{Willott2005}
{Willott}, C.~J., {Percival}, W.~J., {McLure}, R.~J., {et~al.} 2005, \apj, 626, 657, \dodoi{10.1086/430168}

\bibitem[{{Yue} {et~al.}(2024){Yue}, {Eilers}, {Simcoe}, {Mackenzie}, {Matthee}, {Kashino}, {Bordoloi}, {Lilly}, \& {Naidu}}]{Yue2024}
{Yue}, M., {Eilers}, A.-C., {Simcoe}, R.~A., {et~al.} 2024, \apj, 966, 176, \dodoi{10.3847/1538-4357/ad3914}

\bibitem[{{Zastrocky} {et~al.}(2024){Zastrocky}, {Brotherton}, {Du}, {McLane}, {Olson}, {Dale}, {Kobulnicky}, {Maithil}, {Nguyen}, {Chick}, {Kasper}, {Hand}, {Adelman}, {Carter}, {Murphree}, {Oeur}, {Roth}, {Schonsberg}, {Caradonna}, {Favro}, {Ferguson}, {Gonzalez}, {Hadding}, {Hagler}, {Rogers}, {Stack}, {Chapman}, {Bao}, {Fang}, {Zhai}, {Yang}, {Chen}, {Bai}, {Fu}, {Liu}, {Yao}, {Peng}, {Songsheng}, {Li}, {Bai}, {Hu}, {Xiao}, {Ho}, \& {Wang}}]{Zastrocky2024}
{Zastrocky}, T.~E., {Brotherton}, M.~S., {Du}, P., {et~al.} 2024, \apjs, 272, 29, \dodoi{10.3847/1538-4365/ad3bad}

\bibitem[{{Zhang} {et~al.}(2023){Zhang}, {Ouchi}, {Gebhardt}, {Liu}, {Harikane}, {Cooper}, {Davis}, {Farrow}, {Gawiser}, {Hill}, {Kollatschny}, {Ono}, {Schneider}, {Finkelstein}, {Gronwall}, {Jogee}, \& {Krumpe}}]{Zhang2023}
{Zhang}, Y., {Ouchi}, M., {Gebhardt}, K., {et~al.} 2023, \apj, 948, 103, \dodoi{10.3847/1538-4357/acc2c2}

\bibitem[{{Zuo} {et~al.}(2020){Zuo}, {Wu}, {Fan}, {Green}, {Yi}, {Schulze}, {Wang}, \& {Bian}}]{Zuo2020}
{Zuo}, W., {Wu}, X.-B., {Fan}, X., {et~al.} 2020, \apj, 896, 40, \dodoi{10.3847/1538-4357/ab91a7}

\end{thebibliography}
\bibliographystyle{aasjournal}



\end{document}